\documentclass[aps,pra,reprint,groupedaddress]{revtex4-1}
\usepackage{graphicx}
\usepackage{xcolor}
\usepackage{amsmath}
\begin{document}

\def\x{\boldsymbol{x}}
\def\r{\boldsymbol{r}}
\def\n{\boldsymbol{n}}
\def\s{s}
\def\dfrac{\displaystyle\frac}
\def\dint{\displaystyle\int}
\def\rot{\nabla\times}
\def\grad{\nabla}

\def\tensmu{\underline{\underline{\mu}}}
\def\tenseps{\underline{\underline{\varepsilon}}}
\def\mur{\mu_r}
\def\epsr{\varepsilon_r}
\def\Mnm{\ensuremath{\mathsf{\mathbf{M}}_{n,m}^{(+)}}}
\def\Nnm{\ensuremath{\mathsf{\mathbf{N}}_{n,m}^{(+)}}}
\def\Xnm{\ensuremath{\mathsf{\mathbf{X}}_{n,m}}}
\def\Ynm{\ensuremath{\mathsf{\mathbf{Y}}_{n,m}}}
\def\Znm{\ensuremath{\mathsf{\mathbf{Z}}_{n,m}}}
\def\fhnm{f^{(h)}_{n,m}}
\def\fenm{f^{(e)}_{n,m}}
\def\bF{\ensuremath{\mathbf{F}}}
\def\bE{\ensuremath{\mathbf{E}}}
\def\bH{\ensuremath{\mathbf{H}}}
\def\bEp{\mathsf{\mathbf{E}}_{\#}}
\def\bEinc{\mathsf{\mathbf{E}}^{\mathsf{inc}}}
\def\bEd{\mathsf{\mathbf{E}}^{\mathsf{d}}}
\def\bEtot{\mathsf{\mathbf{E}}^{\mathsf{tot}}}
\def\ofx{(\mathbf{x})}
\def\ofkr{(k\mathbf{r})}
\def\ofr{(\mathbf{r})}
\def\oftp{(\theta,\varphi)}
\definecolor{cguill}{rgb}{.7,.7,.7}
\definecolor{cblack}{rgb}{0,0,0}
\def\guill{\color{black}}
\def\black{\color{cblack}}

%

\title{Design of metallic nanoparticles gratings for 
filtering properties in the visible spectrum}

\author{Y. Br{\^u}l{\'e}}
\affiliation{CNRS, Aix-Marseille Université, Centrale Marseille, Institut Fresnel UMR 7249, 13013 Marseille, France}
\author{G. Dem{\'e}sy}
\affiliation{CNRS, Aix-Marseille Université, Centrale Marseille, Institut Fresnel UMR 7249, 13013 Marseille, France}
\author{A.-L. Fehrembach}
\affiliation{CNRS, Aix-Marseille Université, Centrale Marseille, Institut Fresnel UMR 7249, 13013 Marseille, France}
\author{B. Gralak}
\affiliation{CNRS, Aix-Marseille Université, Centrale Marseille, Institut Fresnel UMR 7249, 13013 Marseille, France}
\affiliation{Corresponding author: boris.gralak@fresnel.fr}
\author{E. Popov}
\affiliation{CNRS, Aix-Marseille Université, Centrale Marseille, Institut Fresnel UMR 7249, 13013 Marseille, France}
\author{G. Tayeb}
\affiliation{CNRS, Aix-Marseille Université, Centrale Marseille, Institut Fresnel UMR 7249, 13013 Marseille, France}
\author{M. Grangier}
\affiliation{IEF, CNRS, Univ Paris-Sud, Université Paris-Saclay, 91405, Orsay, France} 
\affiliation{PSA Peugeot Citroën, Direction Scientifique, Centre technique de Vélizy, route de Gisy, 
Vélizy-Villacoublay, F-78140, France}
\author{D. Barat}
\affiliation{PSA Peugeot Citroën, Direction Scientifique, Centre technique de Vélizy, route de Gisy, 
Vélizy-Villacoublay, F-78140, France}
\author{H. Bertin}
\affiliation{IEF, CNRS, Univ Paris-Sud, Université Paris-Saclay, 91405, Orsay, France} 
\author{P. Gogol}
\affiliation{IEF, CNRS, Univ Paris-Sud, Université Paris-Saclay, 91405, Orsay, France} 
\author{B. Dagens}
\affiliation{IEF, CNRS, Univ Paris-Sud, Université Paris-Saclay, 91405, Orsay, France} 

%

\date{\today}

%

\begin{abstract}
Plasmonic resonances in metallic nanoparticles are exploited to create efficient optical filtering functions.
A Finite Element Method is used to model metallic nanoparticles gratings. The accuracy of this method is 
shown by comparing numerical results with measurements on a two-dimensional grating of gold 
nanocylinders with elliptic cross section. Then a parametric analysis is performed in order to design efficient 
filters with polarization dependent properties together with high transparency over the visible range.
The behavior of nanoparticle gratings is also modelled using the Maxwell-Garnett homogenization theory and 
analyzed by comparison with the diffraction by a single nanoparticle. 
The proposed structures are intended to be included in optical systems which could find innovative applications. 
\end{abstract}


\maketitle

\noindent
\section{Introduction}
Metallic nanoparticles supporting plasmonic resonances can be used to design 
optical functions, for instance filtering properties, 
or chemical and bio-sensing \cite{Perney06,Yih06}. 
Such systems offer several advantages, like the possibility to control the 
operating wavelength with the nature of the metal, the size of the particles, the
polarization of the illumination... 
In this paper we focus on the design of spectral filtering property in reflectivity, 
i.e. frequency selective mirror property. In particular, challenges lie
in obtaining a narrow reflexion peak while keeping absorption losses as low as possible 
together with a transmission level averaged over the visible range as high as possible.

Reflective color filters based on plasmonic resonators have already been proposed in 
the literature. In \cite{kumar2012printing}, different arrays of metallic nanodisks are deposited on a 
back reflector made of silver and gold to print a color image at the optical diffraction 
limit. Dense silver nanorods arrays are used in \cite{junahuang2013reflective} to 
create reflective color filters at specific wavelengths depending on the arrays geometry.
Nevertheless, in these propositions, the overall transmission remains low. Core-shell 
nanoparticles, made of silica and silver and embedded into a polymer matrix, 
have been proposed in \cite{hsu2014transparent} to make transparent displays based on resonant nanoparticle
scattering. The considered filters have reflexion properties at defined wavelengths together 
with globally transparency over the visible spectrum. However, the latter structure, based 
on the scattering phenomenon, does not work as a mirror, but as a display with 
virtual image located in the reflecting structure.
High resolution color transmission filtering and spectral imaging have also been achieved 
\cite{xu2010plasmonic}, where plasmonic nanoresonators, formed by subwavelength metal-insulator-metal 
stack arrays, allow efficient manipulation of light and control of the transmission spectra.
Also, highly transmissive plasmonic substractive color filters are proposed in 
\cite{zeng2013ultrathin} with a single optically-thick nanostructured metal layer 
thanks to the counter-intuitive phenomenon of extraordinary transmission.

In this paper, the main objective is to design metallic nanoparticles gratings 
in order to obtain filtering property for wavelengths in visible range together 
with relatively low absorption. First, we present the formulation of the Finite Element 
Method (FEM) \cite{demesy2010all} chosen to model the considered structures, 
whose accuracy has been checked using comparison with the Fourier Modal Method \cite{LiFMM}. 
The relevance of this numerical tool is shown with a comparison between 
full three-dimensional numerical calculations and measurements on fabricated samples.  
The numerical tool is then exploited to perform a parametric analysis of the influence of
geometric parameters and polarization on the filtering properties and absorption. 
Next, this parametric analysis is used to propose optimized filtering systems with 
high level of transparency in average on the visible spectrum. In addition, different 
models are proposed to analyze the behavior of the nanoparticles gratings.
An analytical homogenization model based on the Maxwell-Garnett approximation 
is faced to the rigorous FEM calculations.
Finally, the multipolar radiation of a single nanoparticle 
is compared to the diffractive properties of a grating of the same nanoparticles.

\begin{figure}[h]
\centering
\fbox{\includegraphics[width=0.38\linewidth]{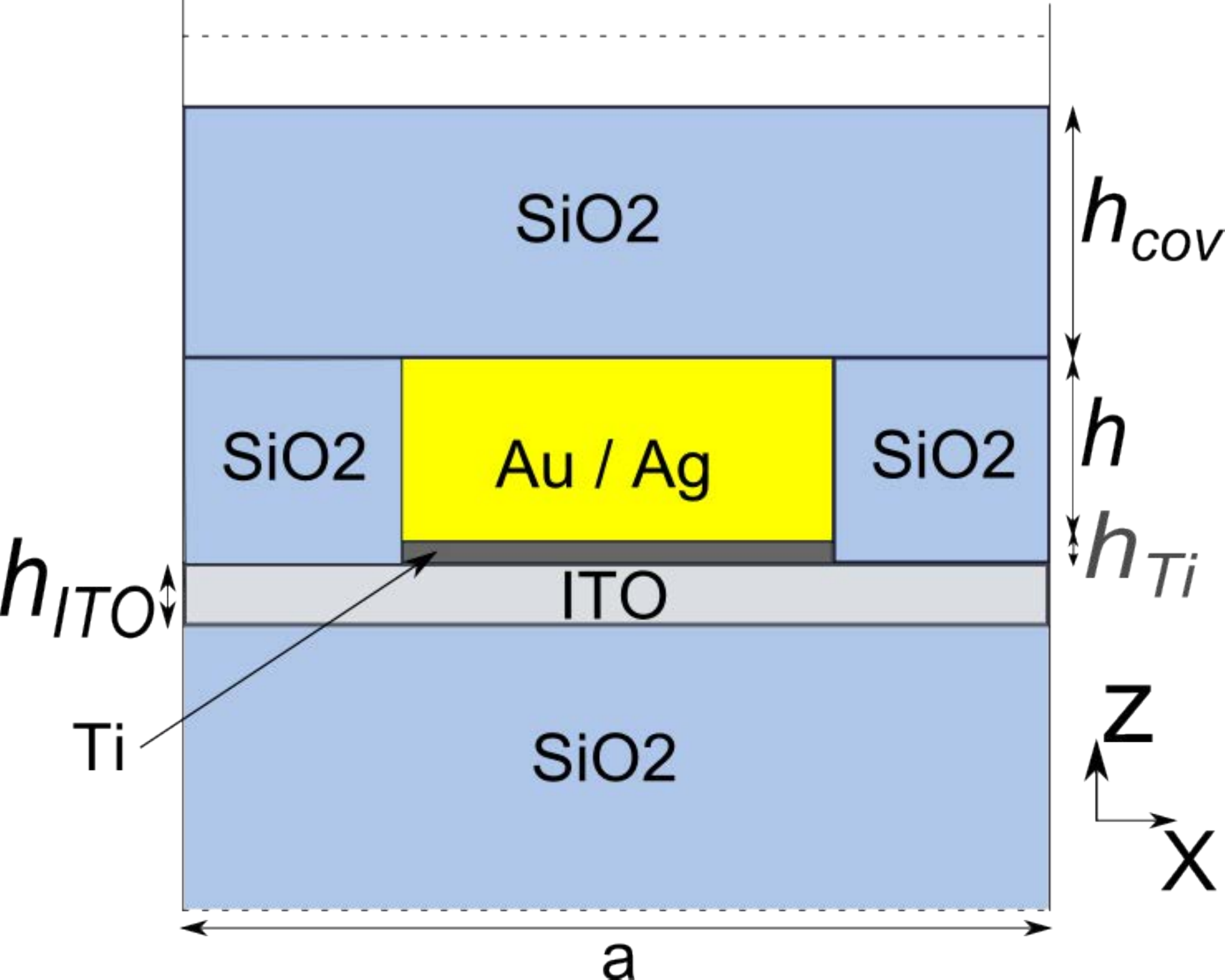}
\includegraphics[width=0.58\linewidth]{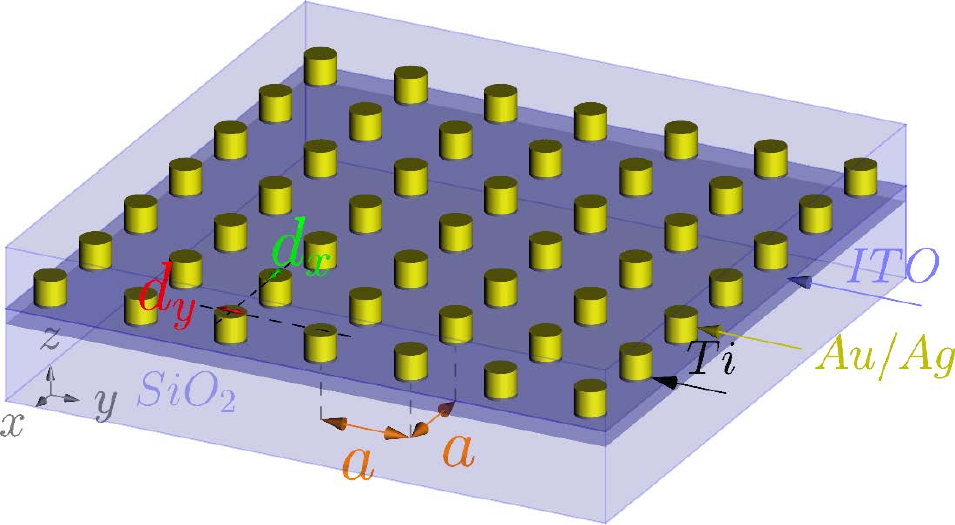}}
\caption{Schemes of the considered nanoparticles gratings.}
\label{fig:Schema}
\end{figure}

The considered gratings are 
defined as two-dimensional square arrays of metallic nanoparticles on a glass 
(SiO$_2$) substrate. The lattice constant of the array along the two directions 
$x$ and $y$ is denoted by $a$. On each node of this array lies a nanocylinder with 
elliptic cross section 
of height $h$, diameters $d_x$ and $d_y$ (along the $x$ and $y$ directions), and made 
of gold (Au) or silver (Ag). Note that, for fabrication requirements, an optional 
Indium Tin Oxide (ITO) thin film layer of height $h_{\text{ITO}}$ may be added on the 
top of the substrate, and a second optional attaching layer of titanium (Ti), of 
height $h_{\text{Ti}}$, may be also added to the bottom of the nanocylinder. 
Finally, nanocylinders grating can be embedded into a SiO$_2$ cover layer of 
height $h_{\text{cov}}$ from the top of the nanocylinder. Figure \ref{fig:Schema} 
presents schemes of the considered structures.
Finally, the refractive indices of the different materials for the simulations are taken from
\cite{weber2002handbook} for SiO$_2$, \cite{johnson1972optical} for gold, 
\cite{palik1998handbook} for silver, 
and \cite{rakic1998optical} for titanium.

\section{Numerical modeling}
\label{sec:Numer}
\subsection{Finite Element Method formulation}
\begin{figure}[h]
\centering
\fbox{\hspace*{5mm}
\includegraphics[width=0.35\linewidth]{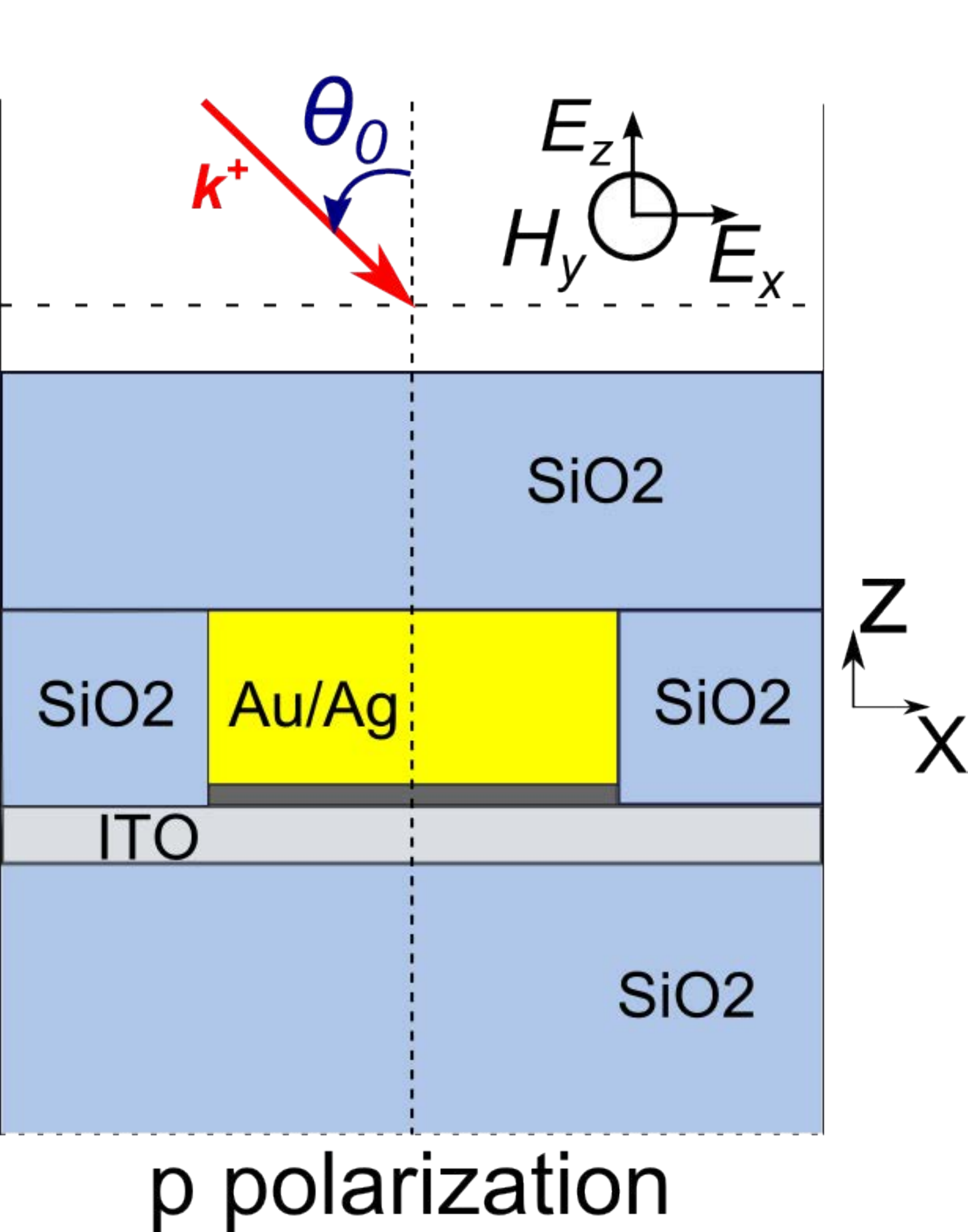}
\hspace*{12mm}
\includegraphics[width=0.35\linewidth]{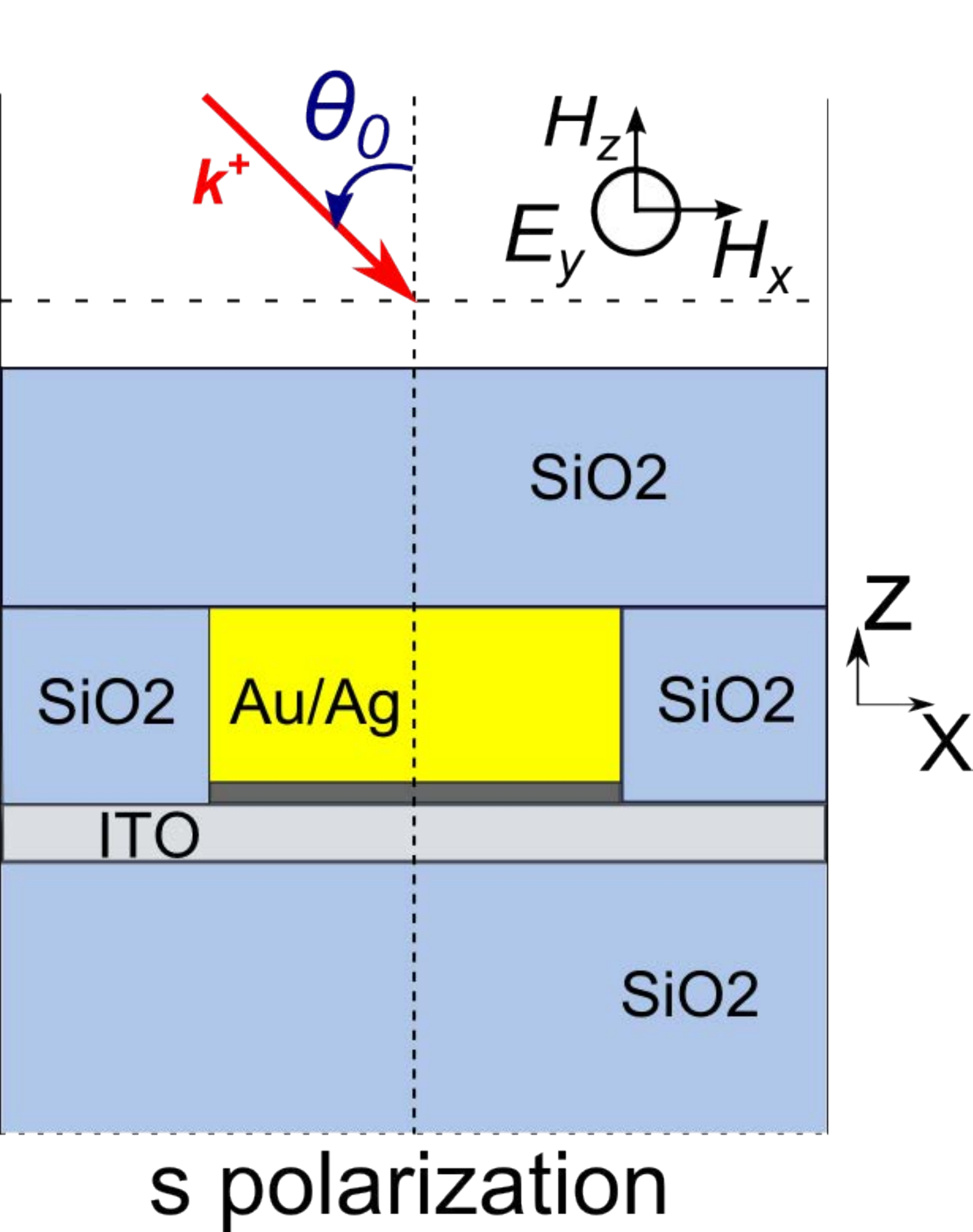}
\hspace*{3mm}
}
\caption{Scheme of the illumination conditions for wavevector $\boldsymbol{k}^+$ lying in 
the $xz$-plane ($\psi_0 = 0$).}
\label{fig:incfield}
\end{figure}

The FEM formulation used in the present paper is described in \cite{demesy2010all}. This 
numerical method allows calculation in the time-harmonic regime of the vector fields 
diffracted by any arbitrarily shaped crossed-grating embedded in a multilayered stack.
The considered grating shown on Fig.\ref{fig:Schema} is placed in an air 
superstrate (incident) medium with permittivity $\epsilon^+ = 1$. The SiO$_2$ substrate 
is considered as semi-infinite with permittivity $\epsilon^-$. All materials of the 
structure are considered to be non magnetic ($\mu_r = 1$). The time-harmonic regime 
with a $\exp[-i \omega t]$ dependence is considered, where the frequency $\omega$ is related
to a wavelength $\lambda = 2\pi c/\omega$ in vacuum, 
$c$ being the speed of light in vacuum. The grating is illuminated with a 
plane wave $\boldsymbol{E}_{0} = \boldsymbol{A}_0\exp[i\,\boldsymbol{k}^+\cdot 
\boldsymbol{x}]$, where the wavevector $\boldsymbol{k}^+$ is defined by 
$\| \boldsymbol{k}^+ \| = {k}^+ = k_0 = \omega /c$ and 
the two angles $\theta_0 \in [0,\pi/2]$ and $\psi_0 \in [0,2\pi]$ 
($\boldsymbol{k}^+$ is in the $xz$-plane if $\psi_0 = 0$ and 
$\boldsymbol{k}^+$ is in the $yz$-plane if $\psi_0 = \pi/2$). This incident plane 
wave is p-polarized if the electric field is inside the incident plane 
(see left panel of Fig. \ref{fig:incfield}) and 
s-polarized if the electric field is perpendicular to the incident plane
(see right panel of Fig. \ref{fig:incfield}).

We want to retrieve the total electromagnetic field $(\boldsymbol{E}\, ,\boldsymbol{H})$ solution 
of the Helmholtz equation
\begin{equation}
\nabla\times \nabla\times \boldsymbol{E}+ k_0^2\mu_0 \epsilon \boldsymbol{E} = 0\, ,
\end{equation}
where the diffracted field defined as $\boldsymbol{E}_{d} = \boldsymbol{E} - \boldsymbol{E}_0$ 
satisfies an outgoing wave condition, and where the field $\boldsymbol{E}$ is quasi-periodic
along the two directions $x$ and $y$.
The vector field $\boldsymbol{E}_{d}$ diffracted by the structure is calculated using 
the solution of an ancillary problem corresponding to the associated multilayered diffraction 
problem (without any diffractive element and with the same condition of illumination). 
This intermediate solution is then used as a known vectorial source term whose support 
is localized inside the diffractive element itself. Hence, the total field $\boldsymbol{E}$ of the 
structure is compound of three fields: i) the diffracted field $\boldsymbol{E}_d$ calculated 
using the FEM; ii) the incident plane wave $\boldsymbol{E}_0$; iii) the plane waves diffracted 
by the multilayer structure (the diffracted field 
associated to the ancillary problem which is compound of the transmitted and 
reflected waves of the multilayer structure).
The total field $\boldsymbol{E}$ having Bloch-boundary conditions along the two directions of 
periodicity, the computation domain is then reduced to a single cell of the 
grating through the set $({k}_x,{k}_y) \equiv ({k}^+_x,{k}^+_y)$ imposed by the incident 
plane-wave. [Note that the two-components wavevector $({k}_x,{k}_y)$ is the same in the whole 
structure with the grating periodicity: hence $({k}^+_x,{k}^+_y)$ is denoted 
by $({k}_x,{k}_y)$.] The implementation of those specific boundary conditions adapted to the 
FEM is described in \cite{nicolet2004modelling}. As mentioned above, the diffracted field 
$\boldsymbol{E}_d$ should also satisfy an outgoing wave condition in the $z$-direction. Thus, 
a set of Perfectly Matched Layer (PML \cite{berenger1994perfectly}) are introduced in order to truncate 
the substrate and superstrate along the $z$ axis. Indeed, the diffracted field that radiates 
from the structure towards the infinite regions decays exponentially inside the implemented 
PML along the $z$ axis. 

Once the values of the fields in the whole structure are obtained, they are used to 
compute a complete energy balance of the bi-periodic grating. It is based on i) 
the calculation of the diffraction efficiencies of each reflected and transmitted 
order along the two directions of periodicity of the gratings through a double 
Rayleigh expansion of the fields, and ii) the calculation of the normalized losses 
of each part of the diffractive (absorptive) element of the cell. This energy 
balance is then used to check the accuracy and self-consistency of the whole 
calculation as the sum of the different components of the energy balance should 
be equal to $1$.

The described method has been implemented into the following FEM free softwares:
Gmsh \cite{geuzaine2009gmsh} as a mesh generator and visualization tool, and
GetDP \cite{dular1998general} as a Finite Element Library. 
Numerous comparisons with the Fourier Modal Method \cite{LiFMM} have been 
performed to check the accuracy of the present FEM method.

\subsection{Comparison with measurements}
\label{sec:Val}
Samples have been fabricated in order to check the relevance of the numerical tool.
We use a commercial SiO$_2$ substrate covered by a 40~nm ITO layer. This ITO layer is used 
to evacuate charges during 
the lithography process. On this ITO thin film, a 4~nm titanium adhesion layer and 
a 30~nm gold film were deposited by electron beam evaporation. The gold nanocylinders 
are then realized by electron beam lithography. Scanning Electron 
Microscopy (SEM) imaging of the realized gratings are then performed and allow 
measurements of dimensions of the realized elliptical nanocylinders.
Three gold nanocylinders gratings with different square periodicities have been fabricated: 
S1 ($a=200$~nm),  S2 ($a=250$~nm) and S3 ($a=300$~nm). 
{\color{black}{
\begin{table}[h]
\begin{center}
\begin{tabular}{|r|c|c|c|}
\hline
& $a$ & $d_x$ & $d_y$ \\
\hline
S1 & $200$~nm & $91$~nm & $129$~nm \\
\hline
S2 & $250$~nm & $87$~nm & $125$~nm \\
\hline
S3 & $300$~nm & $84$~nm & $122$~nm \\
\hline
\end{tabular}
\caption{Dimensions of cylinders of the three fabricated structures.}
\label{tab:S123}
\end{center}
\end{table} 
The elliptical diameters 
of nanocylinders for the three fabricated structures S1--S2--S3 are 
approximately $d_x \sim 80$~nm and $d_y \sim 120$~nm (see table \ref{tab:S123}
for the exact dimensions).}}
Figure 3 shows a scheme and a SEM imaging of one of these gratings.
\begin{figure}[h]
\centering
\fbox{\includegraphics[width=0.42\linewidth]{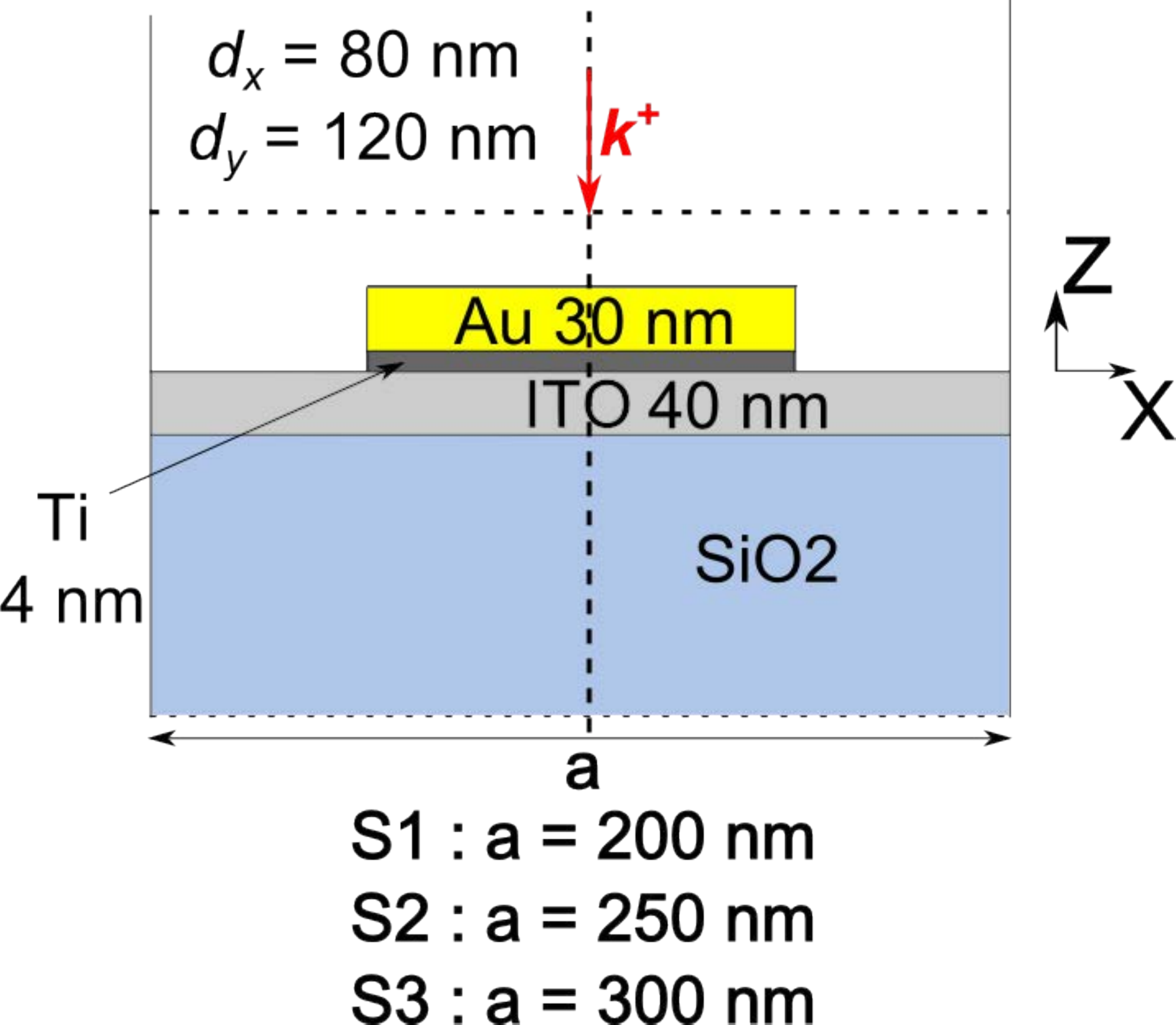}
\includegraphics[width=0.54\linewidth]{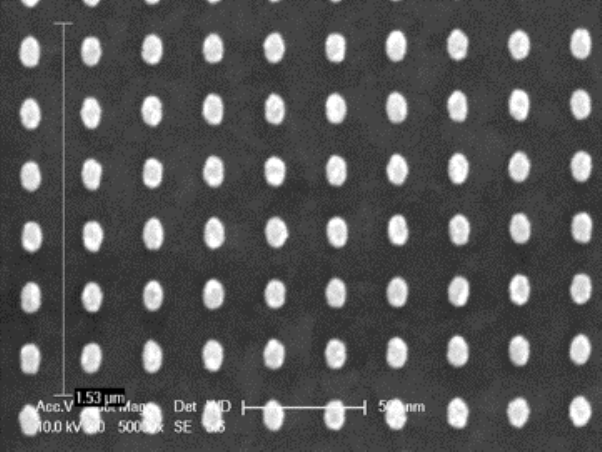} }
\caption{Scheme and SEM imaging of the fabricated structures S1--S2--S3.}
\label{fig:SEM}
\end{figure}

The optical characterization of the samples is based on the measurement of 
transmission. A polychromatic source (Xe lamp) illuminates the 
samples through a monochromator and a polarizer. The source illuminating the samples 
is then almost monochromatic (with a wavelength from 400~nm to 1100~nm) and is linearly 
polarized. 
The transmission measurement is then performed with a home-made angulo-spectral 
optical reflectivity and transmissivity bench. Notice that this transmission measurement 
is normalized with respect to transmission for the bare SiO$_2$ substrate.

Transmission measurements  have been performed for an 
incident beam in normal incidence with electric field polarized along the 
$y$-direction (s-polarisation) and the $x$-direction (p-polarisation) for wavelengths 
ranging from $500$~nm to $1100$~nm. These measurements are compared with the 
simulation results, where the gratings transmission efficiencies have been also 
normalized by the transmission of the bare SiO$_2$ substrate. Notice that the 
Au refractive index considered in these simulations for comparison is 
determined from measurements on the samples (there is a small difference from datas in 
\cite{johnson1972optical}).
\begin{figure}[h]
\centering
\fbox{
\includegraphics[width=0.47\linewidth]{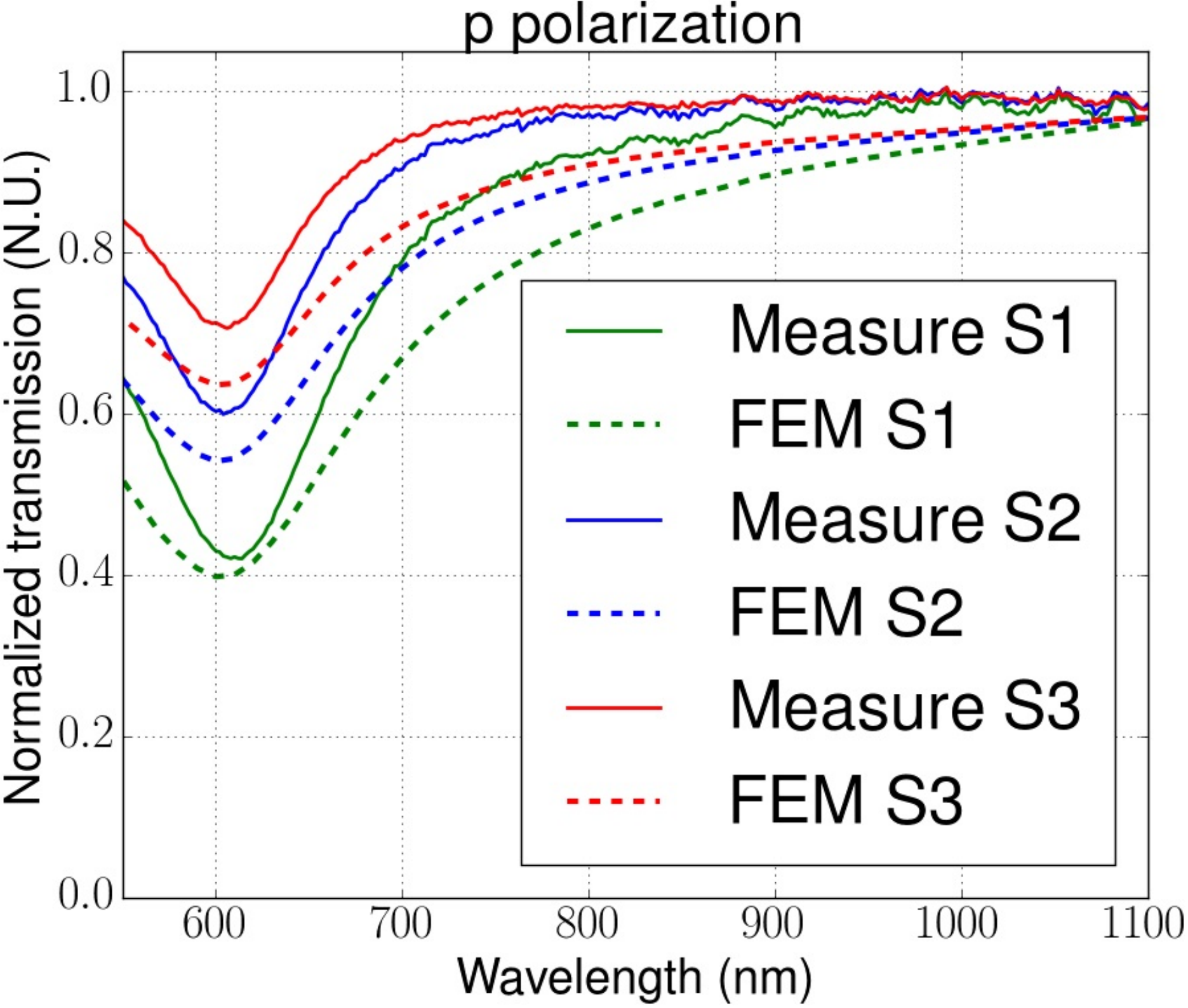}
\includegraphics[width=0.48\linewidth]{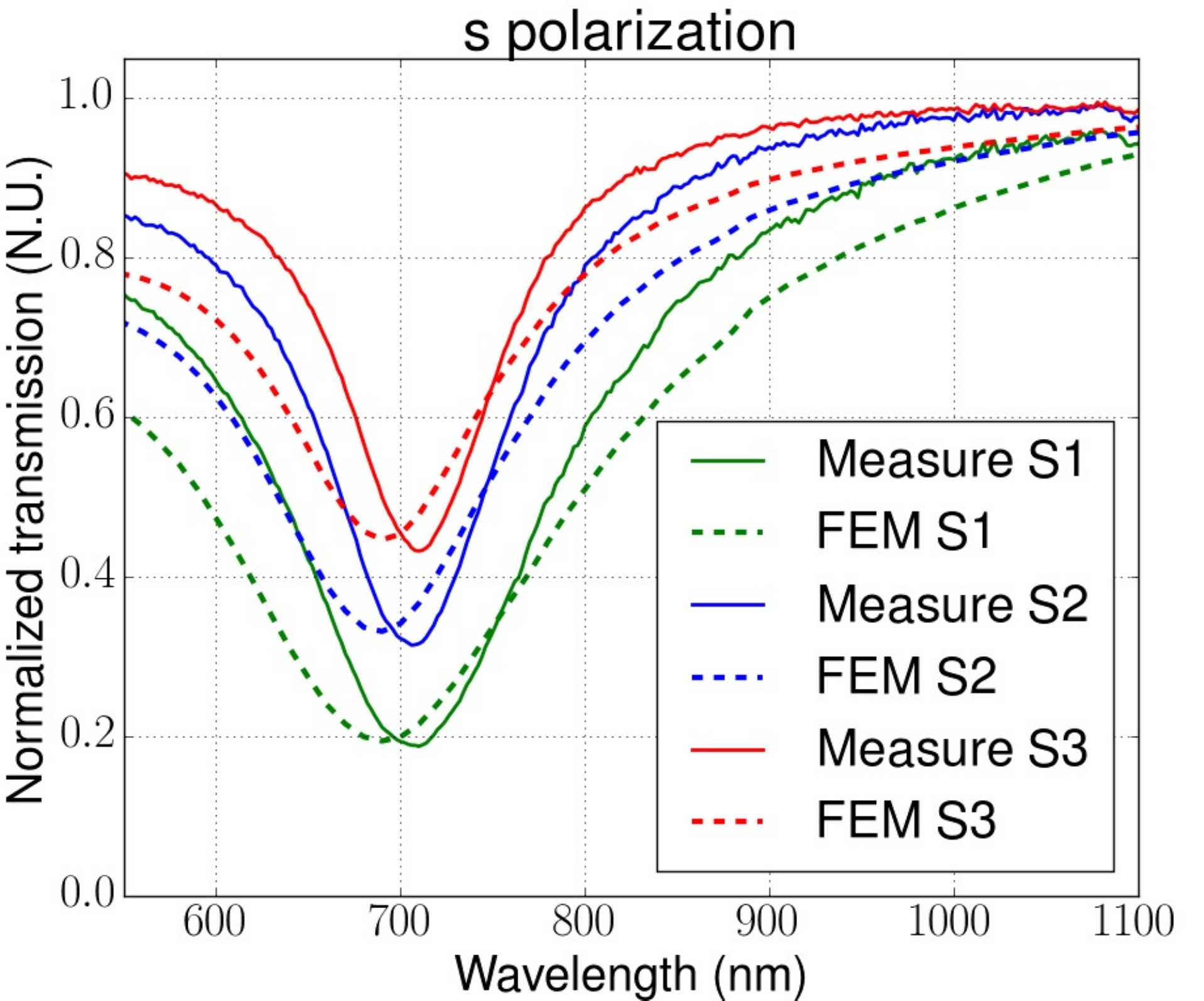}}
\caption{Comparison between the FEM results (dashed) and the measurements (line) in 
transmission for the three fabricated samples S1--S2--S3 .}
\label{SimuMesure}
\end{figure}
{\color{black}{Figure \ref{SimuMesure} shows excellent qualitative and quantitative agreement, and thus the 
relevance of the numerical tool employed to design plasmonic filters. 
The small remaining differences between the measurements 
and calculations can be attributed to the deviations in the shape of the 
fabricated cylinders, for example slanted walls or rounded corners.}}

\section{Parametric analysis and design of 
filtering properties}
\label{sec:Design}
In this section, the influence of the different parameters of the system 
(nature of the metal, geometric parameters and illumination polarization) 
is analyzed in the case of an illumination with oblique incidence 
($\theta_0 = 45^\circ$). This analysis is used to design an optimized structure 
for filtering properties. Here, the challenges are to obtain a choice of resonances 
in the visible spectrum with a minimum of optical losses or a maximum of 
averaged transmission. 

From results presented in the previous section, it appears that structures of Au nanocylinders 
lead to resonances around 700~nm. Hence, Ag nanocylinder are considered in subsection 3.A 
to address the whole visible spectrum for the reflectivity peaks. Next, in 
subsection 3.B, the presence of absorption is analyzed starting from the reference structures 
S1--S2--S3 of Au nanoparticles. Finally, in subsection 3.C, optimized structures are 
proposed.

\subsection{Influence of the cylinder diameters 
and of the incident polarization on the resonance wavelength.}

\begin{figure}[h]
\centering
\fbox{\hspace*{7mm}\includegraphics[width=0.8\linewidth]{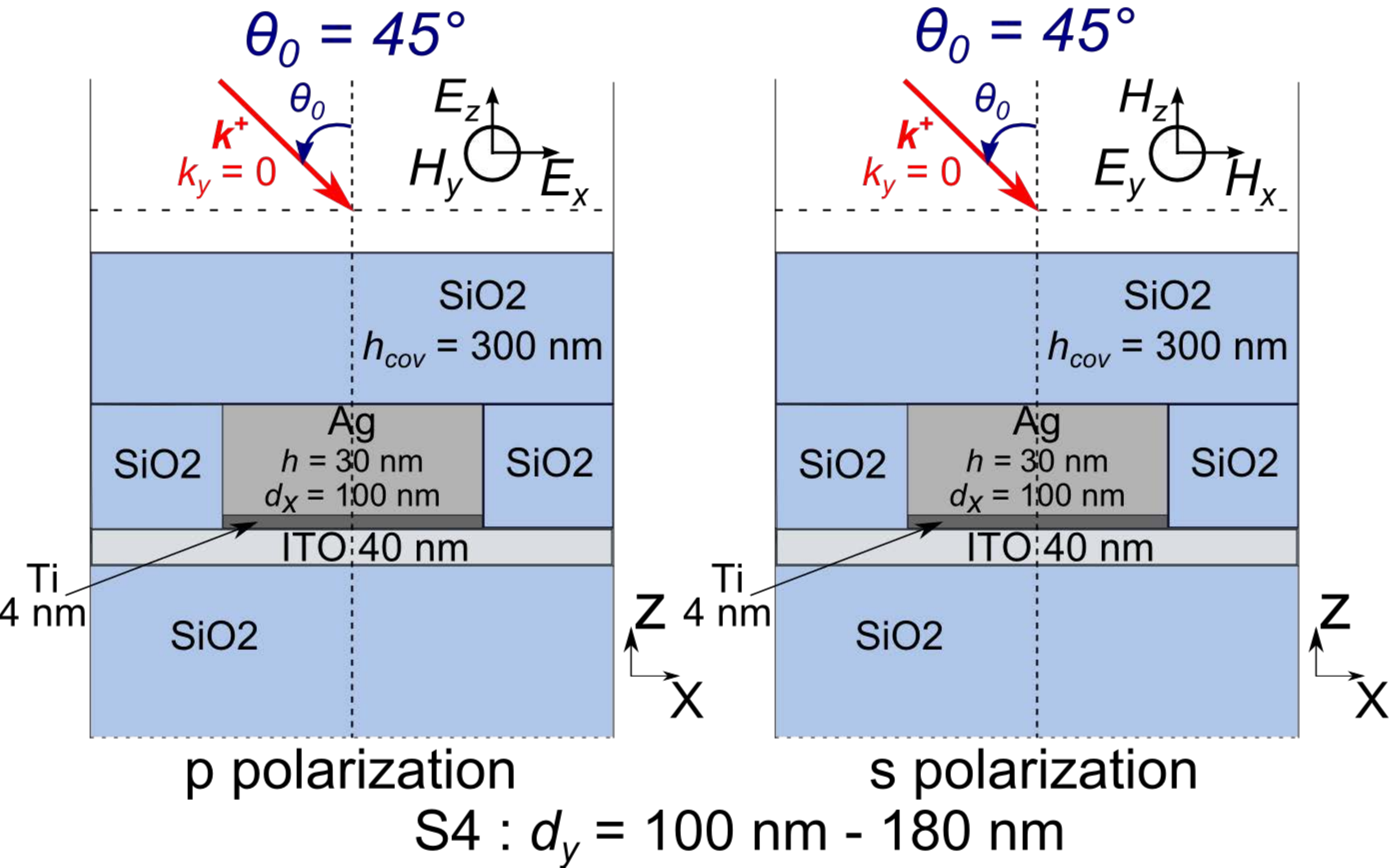}\hspace*{7mm}}
\caption{Scheme of the structure S4 for the considered conditions of illumination.}
\label{fig:S4}
\end{figure}

The geometries considered in this subsection 3.A are similar to the ones 
of subsection 2.B (see Fig. \ref{fig:S4}) with the same ITO and Ti layers. 
The differences are: a $300$~nm SiO$_2$ cover layer
in order to prevent the oxidation of the Ag nanocylinders; square periodicity of the 
gratings fixed to $a = 200$~nm; Ag elliptical nanocylinders with axes  
$d_x$ fixed to $100$~nm and $d_y$ varying from $100$~nm to $180$~nm with a 
$20$~nm pitch (structures of type S4). Here, Ag nanocylinders have been 
introduced in order to address the whole visible spectrum with plasmonic 
resonances. 

\begin{figure}[h]
\centering
\fbox{\includegraphics[width=0.48\linewidth]{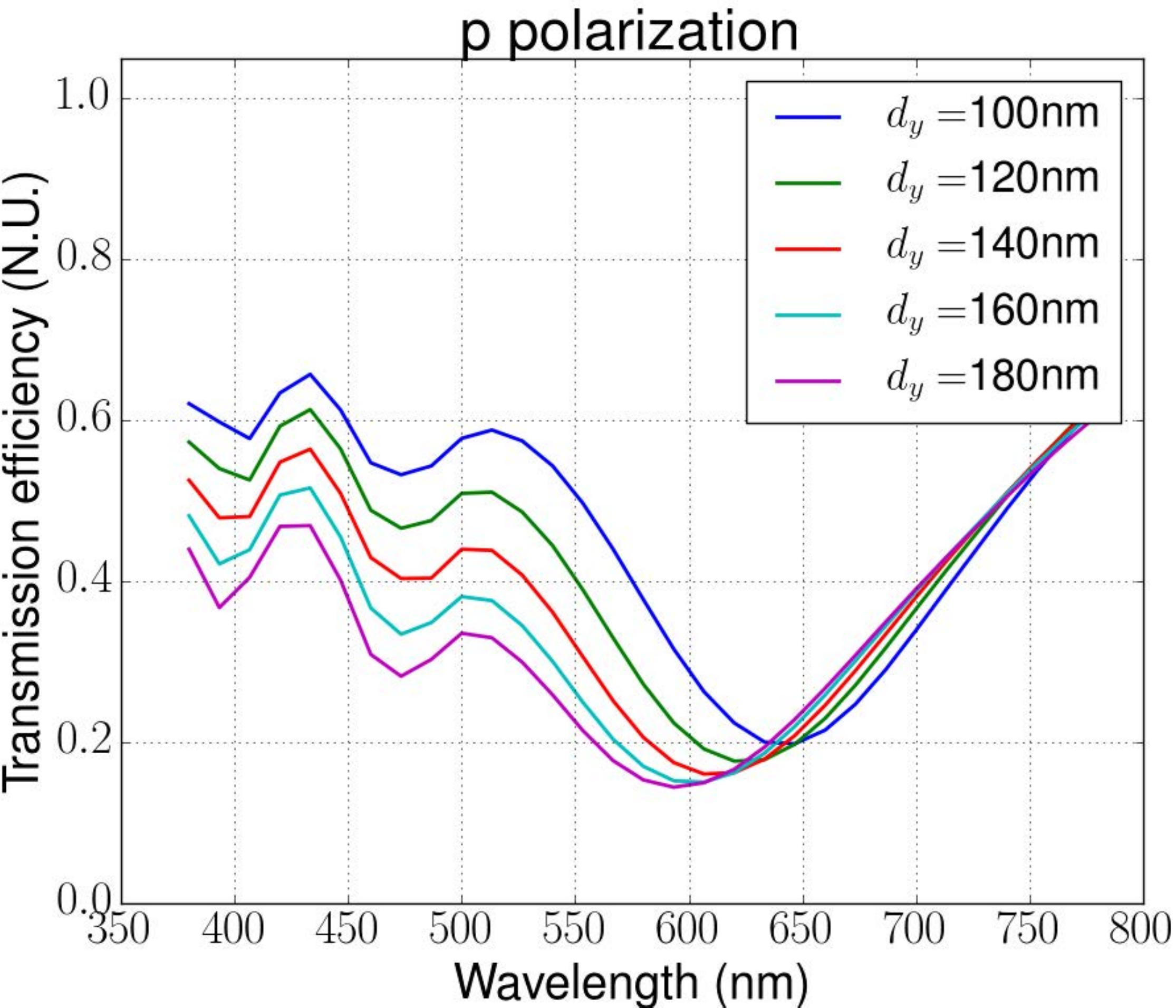}
\includegraphics[width=0.48\linewidth]{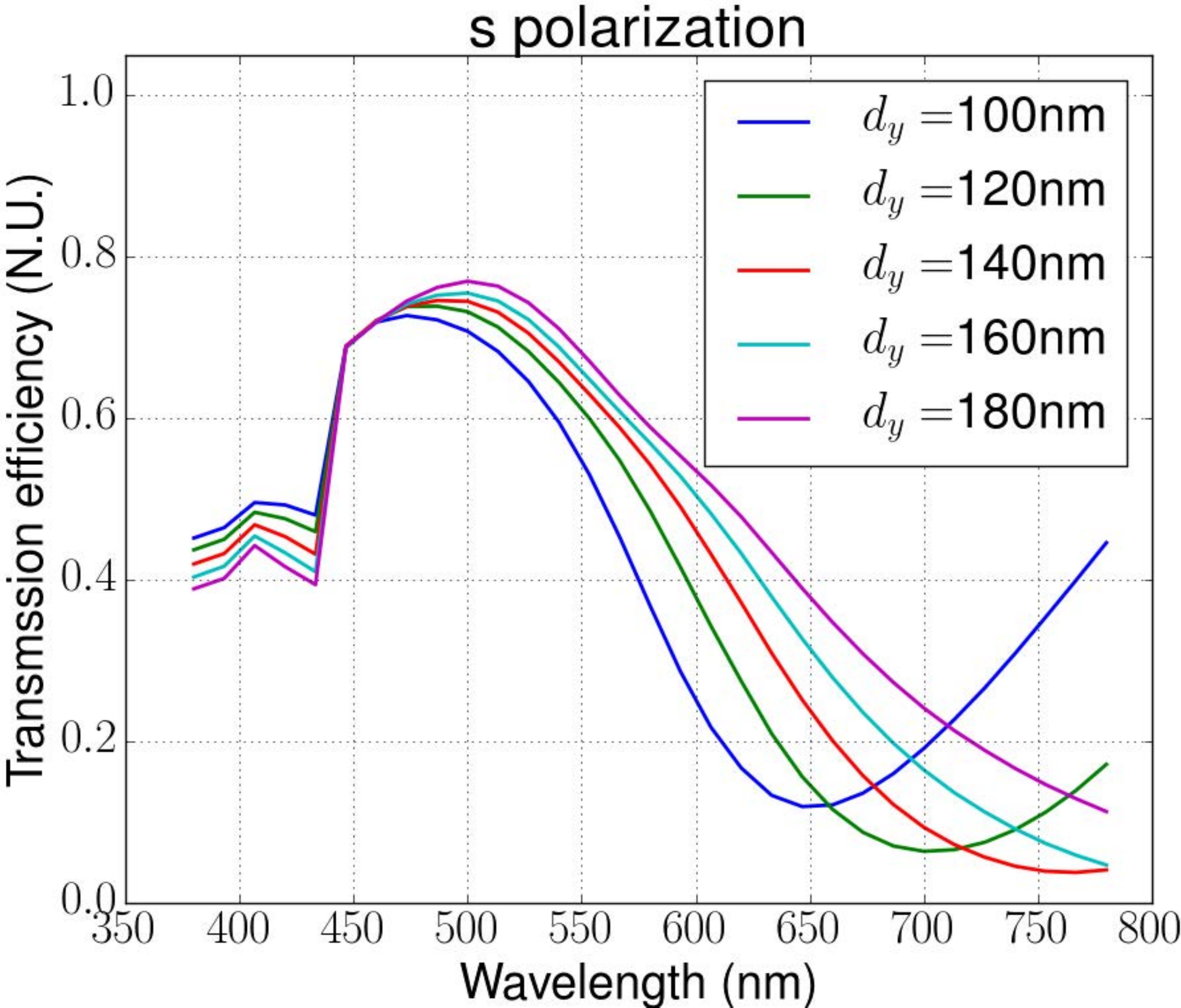}}
\caption{Efficiency in the transmission specular order for the group of structures S4.}
\label{fig:TransmissionDiameterY}
\end{figure}

Simulations are performed for a plane wave for a wavevector lying in the 
$xz$-plane ($k_y = 0$), and the angle of incidence with respect to the normal of the 
substrate is $\theta_0 = 45^\circ$. 
Both polarizations are considered (see Fig. \ref{fig:S4}).
The simulations are performed for 
wavelengths spanning the visible range (380 to 780~nm). Fig. \ref{fig:TransmissionDiameterY} 
shows the transmission efficiency in the specular order for the various $d_y$ diameters, 
for p and s polarizations. 
\begin{figure}[b]
\centering
\fbox{\includegraphics[width=0.48\linewidth]{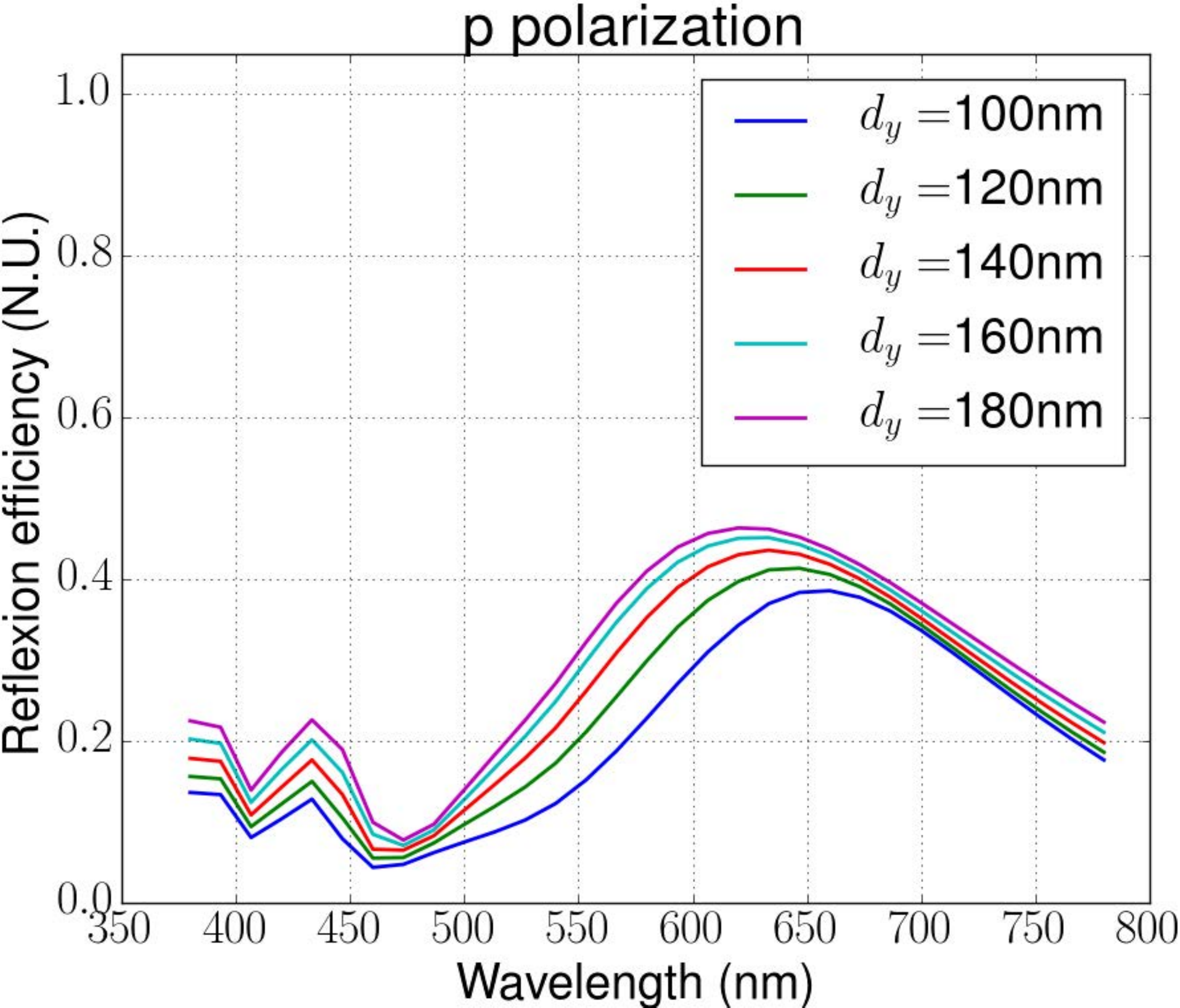}
\includegraphics[width=0.48\linewidth]{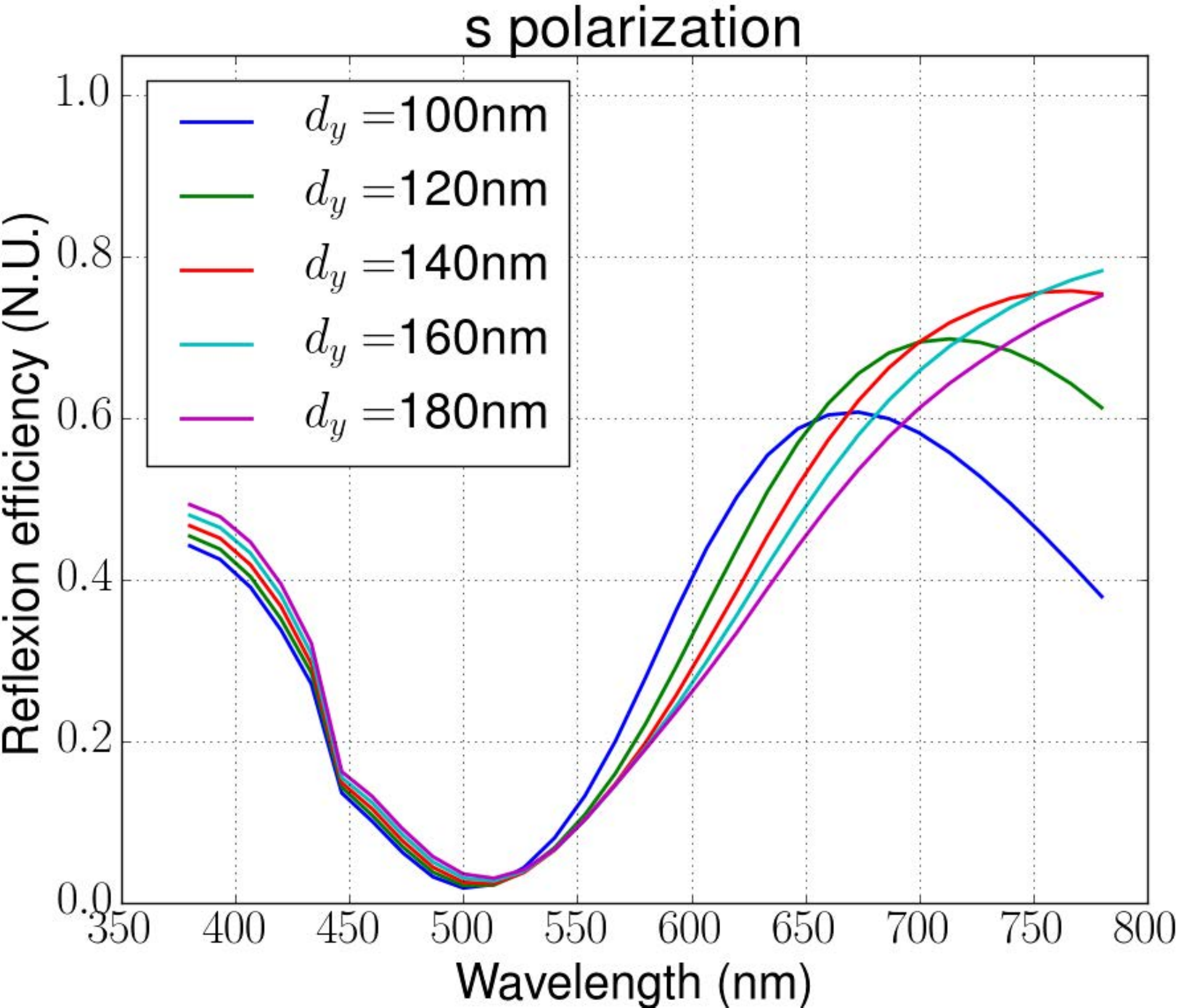}}
\caption{Efficiency in the reflexion specular order for the group of structures S4.}
\label{fig:ReflexionDiameterY}
\end{figure}
The corresponding efficiencies in the reflexion specular 
order are reported in Fig. \ref{fig:ReflexionDiameterY}.
In the present case of an incident plane wave with wavevector lying in the $xz$-plane 
($k_y = 0$), when the plane wave is p-polarized, the electric field ``sees'' the small axis $d_x$ 
of the elliptical nanocylinder while, when the plane wave is s-polarized, the 
electric field ``sees'' the large axis $d_y$. One can observe in Figs. 
\ref{fig:TransmissionDiameterY} and \ref{fig:ReflexionDiameterY} that the resonance 
occuring for p polarization is at lower wavelength and less efficient that the one 
that occurs for s polarization. The more the incident electric field meets a high 
quantity of metal, the more the resonance is pronounced and redshifted. Also, in the 
two cases of polarizations, increasing $d_y$ broadens the reflexion peaks, which is coherent 
with an increasing quantity of metal. In the same time while 
increasing $d_y$, one can observe that the reflexion peak in p polarization is 
blueshifted while the one in s polarization is redshifted. Hence the two resonances 
in the two polarizations are not totally independent, but the general tendencies described 
above remain valid. These effects may allow us to manipulate the resonances observed in those 
elliptical nanocylinders gratings and to select wavelengths for which the reflexion peaks 
occur depending on the plane of incidence of the waves and their polarization. 
Hence the first challenge to obtain a resonance at the desired wavelength in the whole visible 
spectrum might be addressed. 

\subsection{Absorption and influence of the geometrical parameters}

The second challenge is to reduce the absorption in order to obtain a high 
transmission when averaged on the whole visible spectrum. The influence of 
the quantity of metal is thus investigated in detail. 

\begin{figure}[h]
\centering
\fbox{\includegraphics[width=0.45\linewidth]{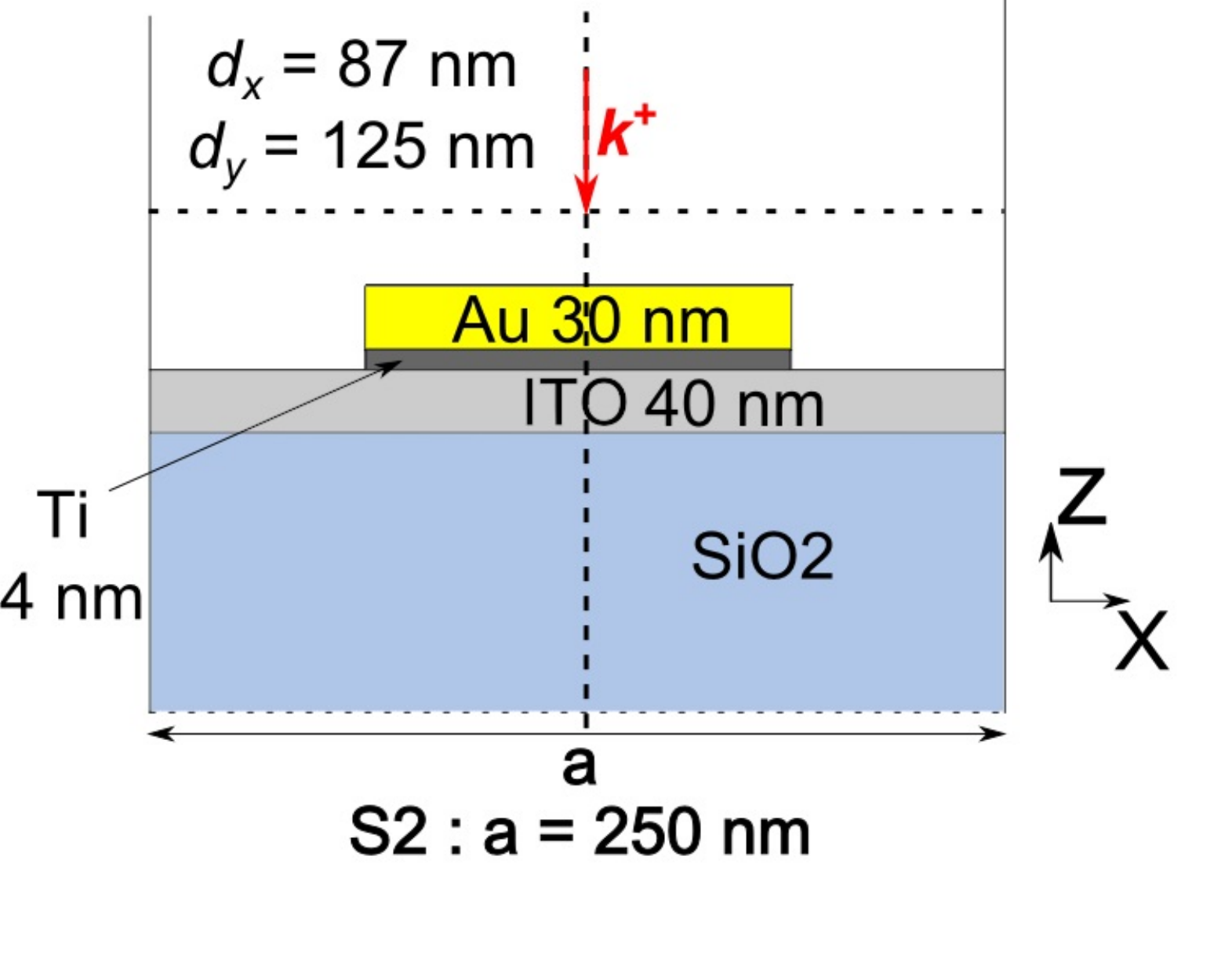}
\includegraphics[width=0.50\linewidth]{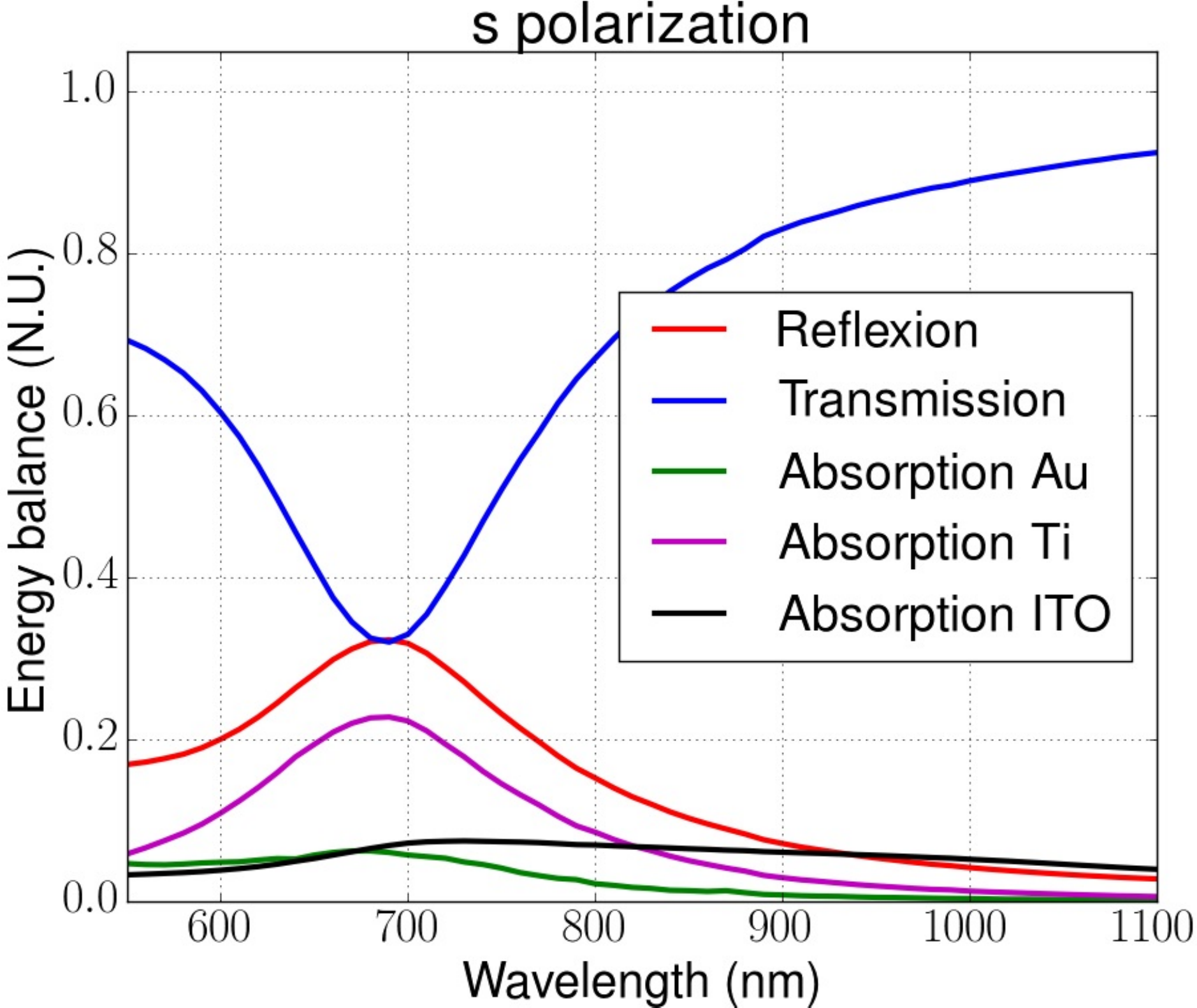}}
\caption{Decomposition of the total energy balance for s-polarization for the 
structure S2.}
\label{DecompAbs}
\end{figure}

As a preliminary, we observe that, from the simulated total energy balance obtained 
for structure S2 (see Fig. \ref{DecompAbs}), losses are important inside the ITO and Ti layers.
Since these materials play no role in the filtering function, suppressing  
ITO and Ti layers should be addressed in order to reduce Joule losses.
From now, the investigation is performed with 
gold nanocylinders gratings directly deposited 
on the SiO$_2$ substrate. 

\begin{figure}[h]
\centering
\fbox{\hspace*{5mm}\includegraphics[width=0.36\linewidth]{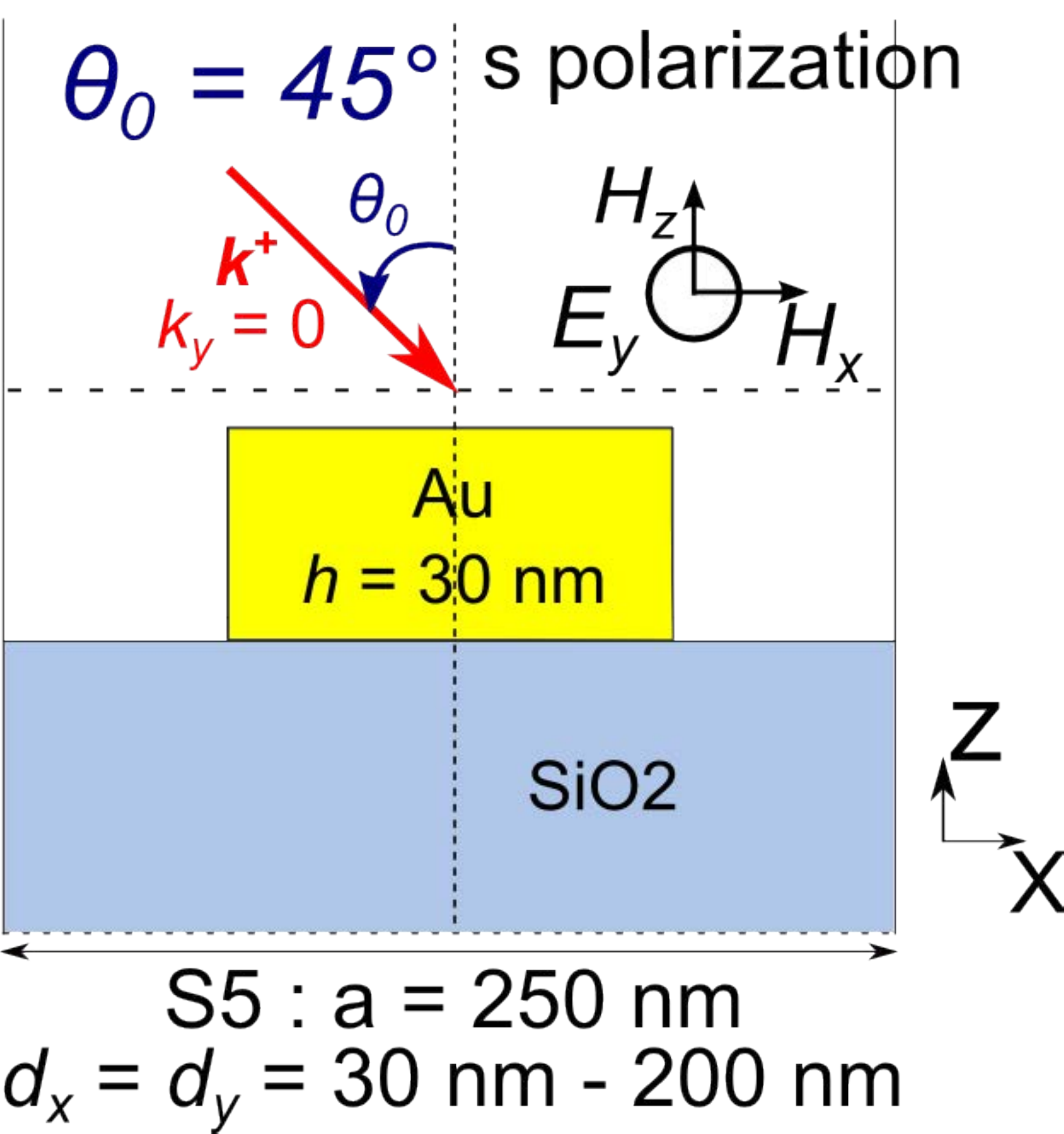}
\hspace*{10mm}
\includegraphics[width=0.36\linewidth]{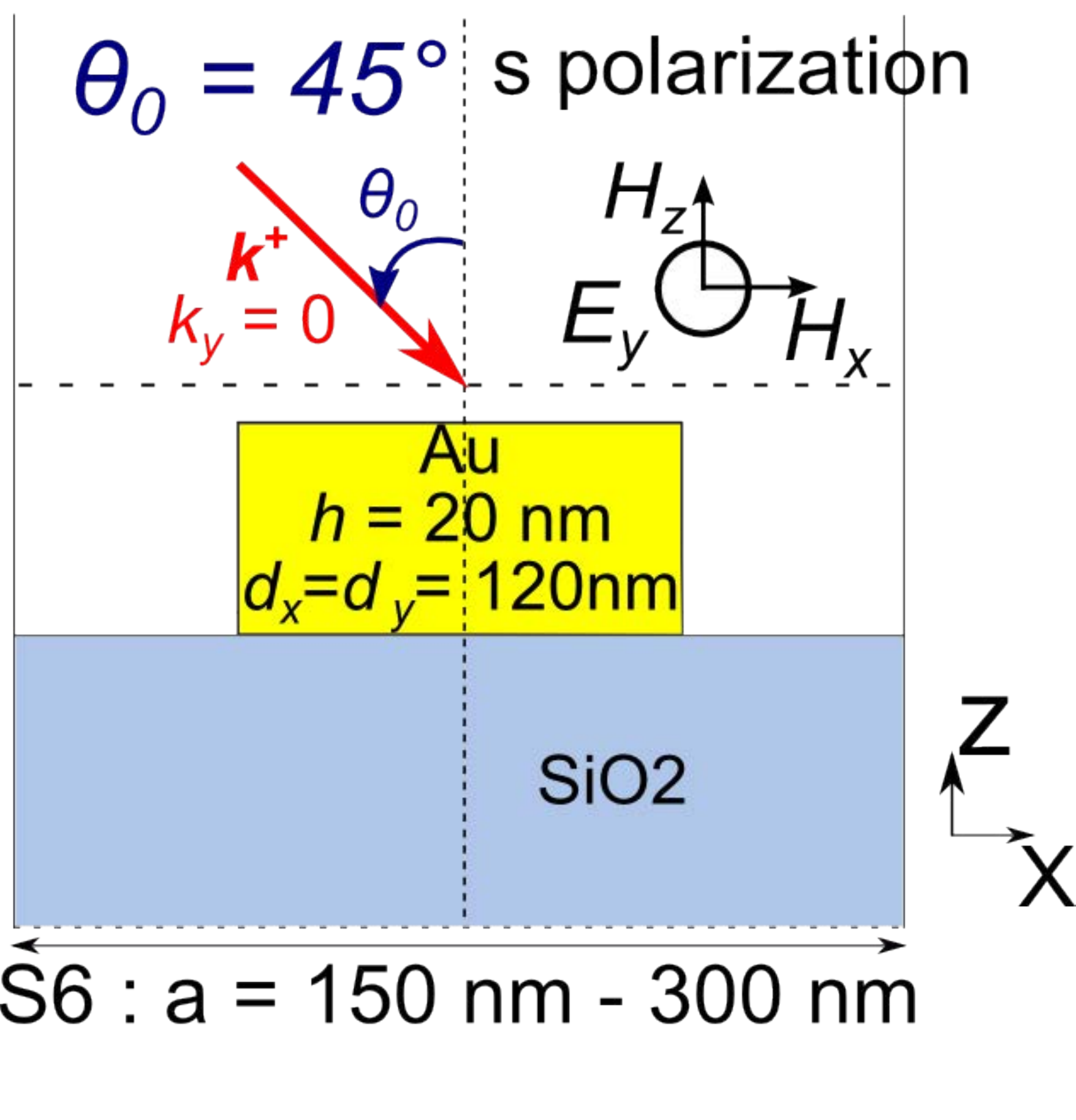}\hspace*{5mm}}
\caption{Scheme of the structures S5 (left panel) and S6 (right panel).}
\label{S5-S6}
\end{figure}
The influence of the quantity of metal is investigated by increasing the filling fraction 
of metal using two groups of structures. For the first group of structures S5 
(left panel of Fig. \ref{S5-S6}), the size of the nanocylinders is increased for a fixed 
lattice constant $a$ and, conversely, the second group of structures S6 
(right panel of Fig. \ref{S5-S6}), the size of the lattice constant $a$ 
is increased for a fixed nanocylinder dimension. Also, to simplify the number of parameters 
of influence, nanocylinders with circular cross section are considered here. The plane wave 
illuminating the structures has wavevector lying in the $xz$-plane ($k_y=0$) with 
an angle of incidence fixed to $\theta_0=45^\circ$, and is s polarized.

First the influence of the nanocylinder diameter on the total energy 
balance spectrum is analyzed. {Notice that the propagative diffraction efficiencies
higher than the $(0,0)$ specular orders are omitted since they are found to be negligible. The 
square pitch of the grating is fixed with $a=250$~nm, the gold nanocylinder height is 
fixed to $h=30$~nm, and their diameters vary from 30 to $200$~nm with $10$~nm pitch 
(group of structures S5). The left panel of Fig. \ref{fig:Reflexion} shows the reflexion
efficiency in the specular order on the visible range ($380$~nm -- $780$~nm) 
as a function of the nanocylinder diameter.
\begin{figure}[h]
\centering
\fbox{\includegraphics[width=0.48\linewidth]{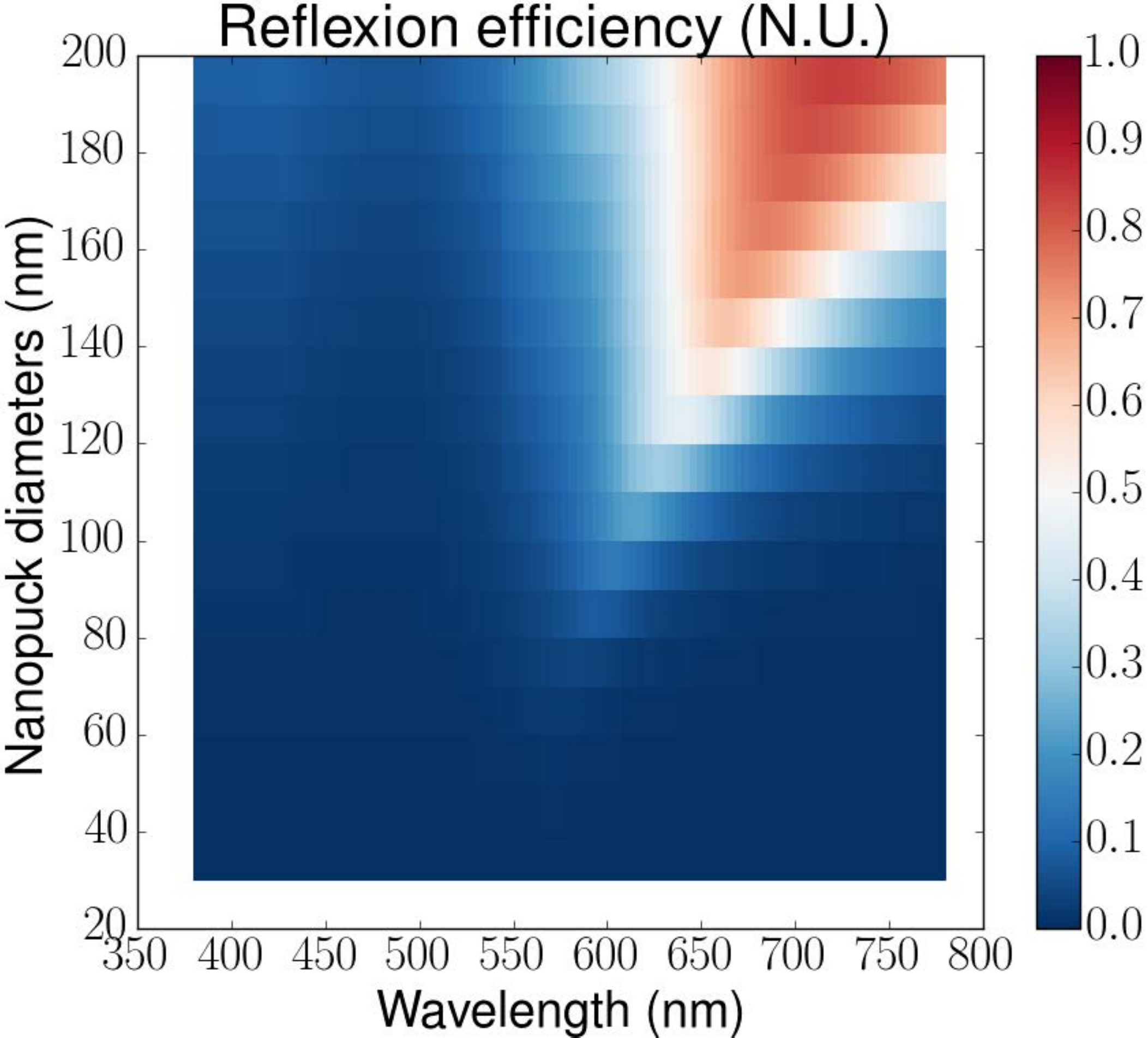}
\includegraphics[width=0.48\linewidth]{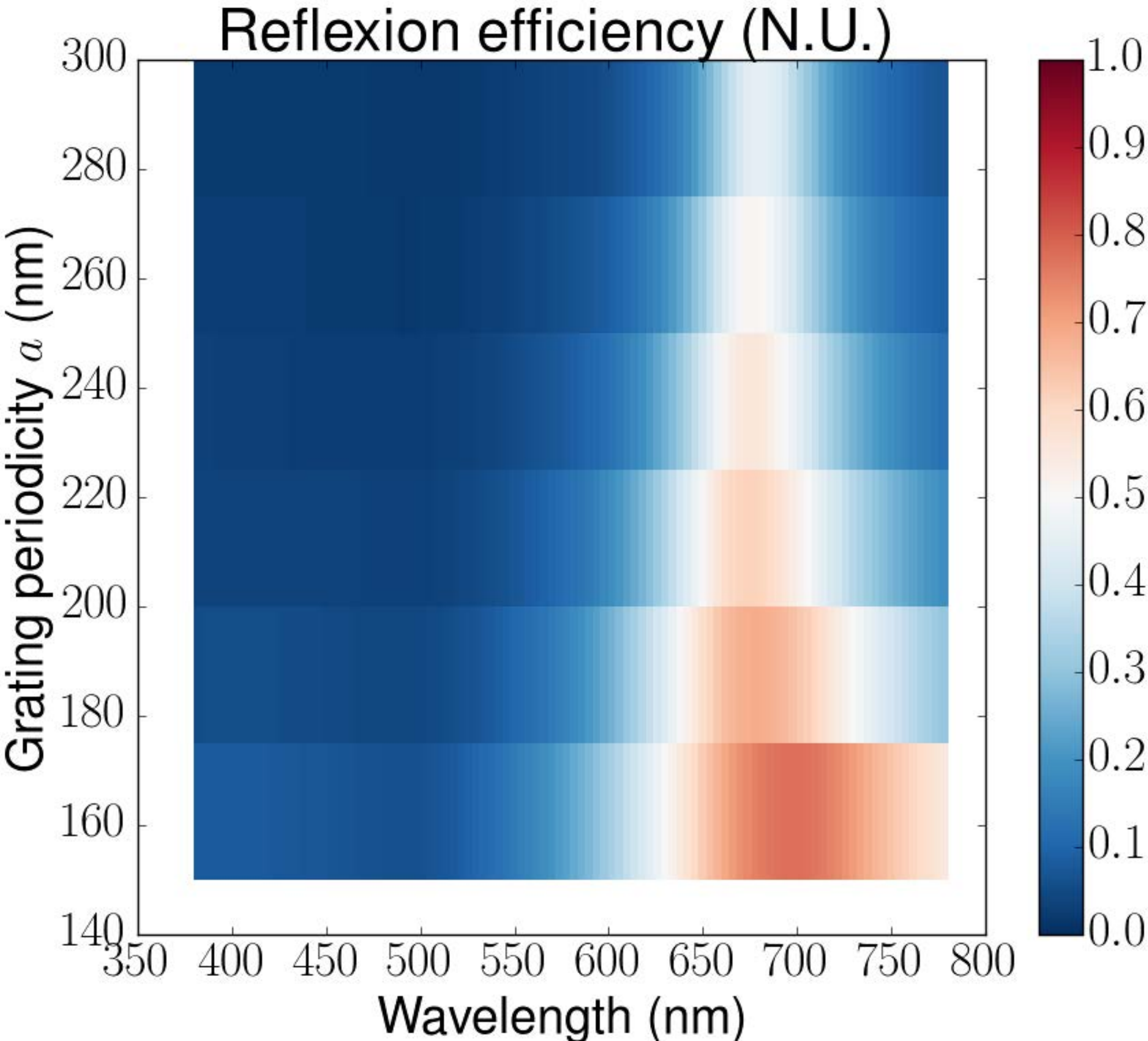}}
\caption{Left panel: Reflexion efficiency in the specular order for structures S5. 
Right panel: Reflexion efficiency in the specular 
order for structures S6.}
\label{fig:Reflexion}
\end{figure}
One can observe that as the nanocylinder diameter increases, the reflexion peak 
intensity is magnified, redshifted and broadened. Again, one can check the possibility to 
choose the operating wavelength of the filter by selecting the appropriate diameter.

Second, the influence of the lattice constant $a$ on the total energy 
balance spectrum is analyzed. The nanocylinder geometry is then fixed with 
a height $h= 20$~nm and a diameter $d_x = d_y = 120$~nm. The square periodicity $a$ 
is now spanning the range (150~nm -- 300~nm) with a 25~nm pitch (group of structures S6). 
The configuration of illumination remains unchanged (incident angle of $45^\circ$, $k_y = 0$, 
s polarization). The right panel of Fig. \ref{fig:Reflexion} presents the spectrum 
of the reflexion efficiency in the specular order over the visible range 
($380$~nm -- $780$~nm) as a function of the lattice constant $a$. 
For fixed nanocylinder dimensions, increasing the grating 
square periodicity leads to slight variations of the central wavelength of the 
reflectivity peaks, which confirms that the resonance frequency is mainly 
governed by the particle diameter. Moreover, the intensity decreases and the peak is 
sharpened with the increase of the gratings periodicity. This result is fully 
consistent with the study of the influence of diameter.

Combination of these two parametric investigations leads to the following conclusions: i) the 
resonance frequency is mainly governed by the nanocylinder dimensions; ii) the ratio 
of the nanocylinder cross section surface over the lattice unit cell surface $a^2$ 
mainly governs the resonance intensity and absorption. 

Before to present optimized structures, it is stressed that, in the configurations
S5 and S6, efficient reflectivity peaks 
occurs for wavelengths around $650$~nm. If the 
structure is embedded into SiO$_2$ (like for structures S2 in section 2), 
then the observed peak in reflexion is redshifted and will occur only in the 
extreme red part of the visible spectrum, around $700$~nm (for structures S2). 
Thus, as conclusion iii), it appears necessary to consider the solution based on Ag 
proposed in the subsection 3.A to blueshift the resonances, since plasma frequency 
of Ag occurs at lower wavelengths. Finally, as pointed out in the preliminaries of this 
subsection 3.B, high absorption appears in ITO and 
Titanium layers (Fig. \ref{DecompAbs}) and thus the optimization should be performed 
for nanocylinders standing directly on SiO$_2$ substrate (conclusion iv).

\subsection{Optimized design with silver elliptical nanocylinders}
\label{sec:Opti}
In this section, we propose two different designs of silver elliptical nanocylinders
gratings embedded in SiO$_2$ leading to peaks in the specular reflectivity in the visible range. 
The elliptical cross section of the nanocylinders allows two different peaks at two different wavelengths 
for a unique structure, depending on the polarization of illumination. The proposed filters possess also 
properties of global transparency on the visible spectrum.

\begin{figure}[h]
\centering
\fbox{\hspace*{4mm}
\includegraphics[width=0.4\linewidth]{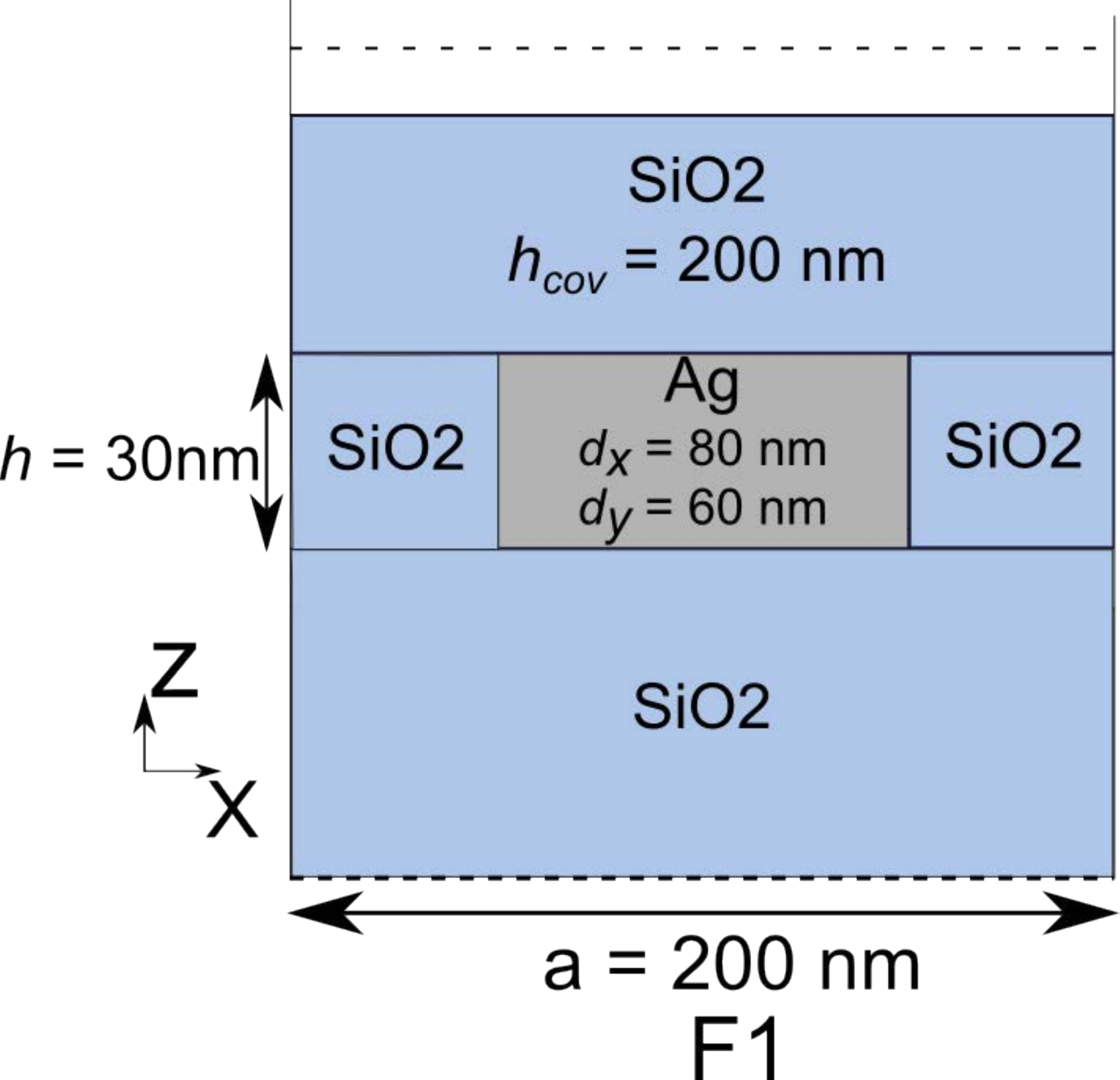}
\hspace*{5mm}
\includegraphics[width=0.4\linewidth]{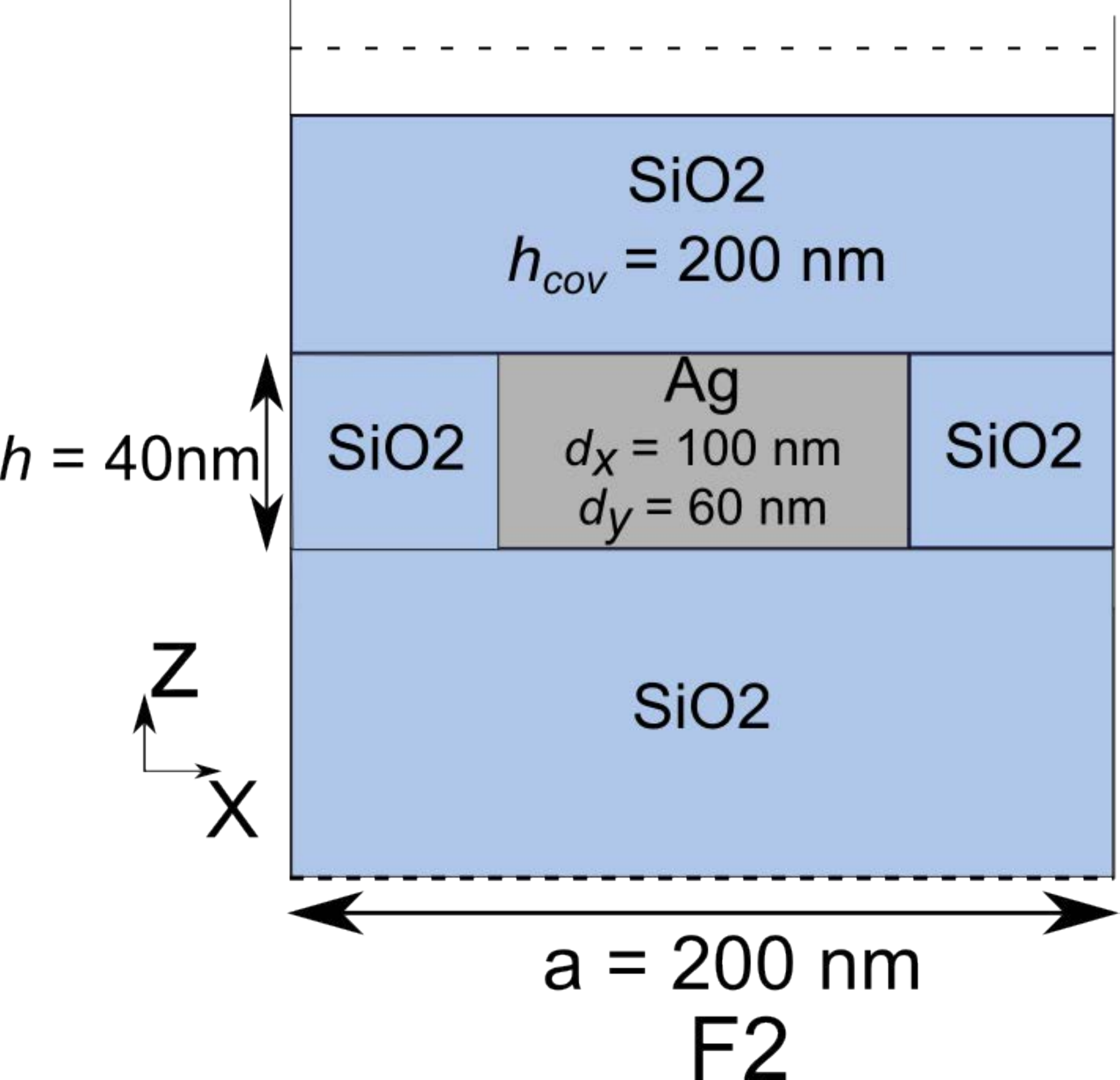}\hspace*{4mm}}
\caption{Schemes of the optimized filters F1 (left panel) and F2 (right panel).}
\label{F1-F2}
\end{figure}

The first proposed filter F1 is constituted of nanocylinders of $30$~nm height (left panel of Fig. 
\ref{F1-F2}). The ellipse diameters 
are $d_x = 80$~nm and $d_y = 60$~nm. The grating is embedded into SiO$_2$ (the substrate is SiO$_2$ 
and a SiO$_2$ cover layer of $200$~nm is added) and the square periodicity $a$ of the grating is fixed to $200$~nm. 
The incident wavevector lies either in the $xz$-plane ($k_y = 0$) or in the $yz$-plane ($k_x = 0$), 
with an angle of incidence fixed to $\theta_0 = 45^\circ$, and both p and s polarizations are considered.
Figure \ref{fig:ROpti590} shows the different 
components of the energy balance for the filtering structure F1.
\begin{figure}[h]
\centering
\fbox{\includegraphics[width=0.48\linewidth]{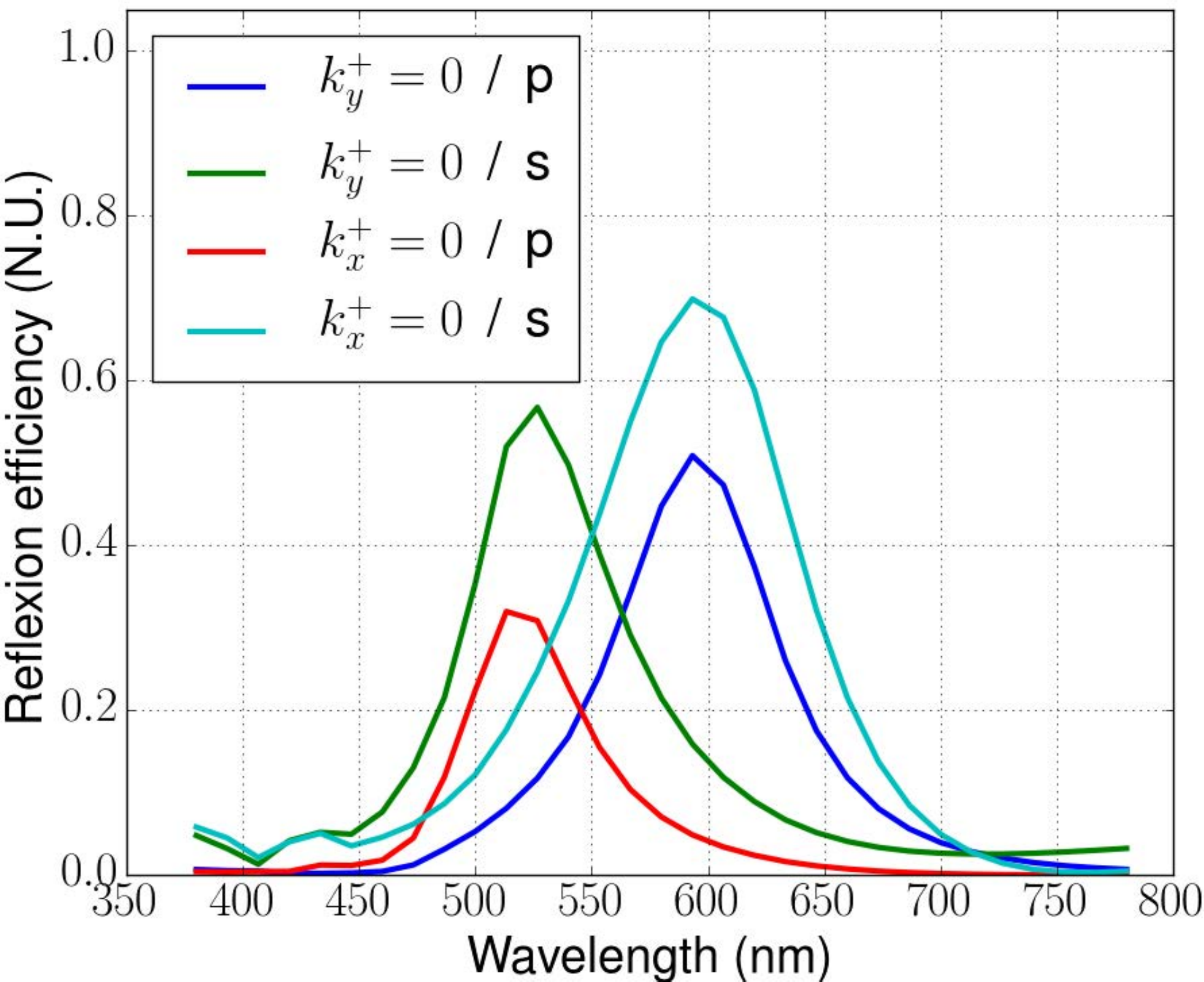}
\includegraphics[width=0.48\linewidth]{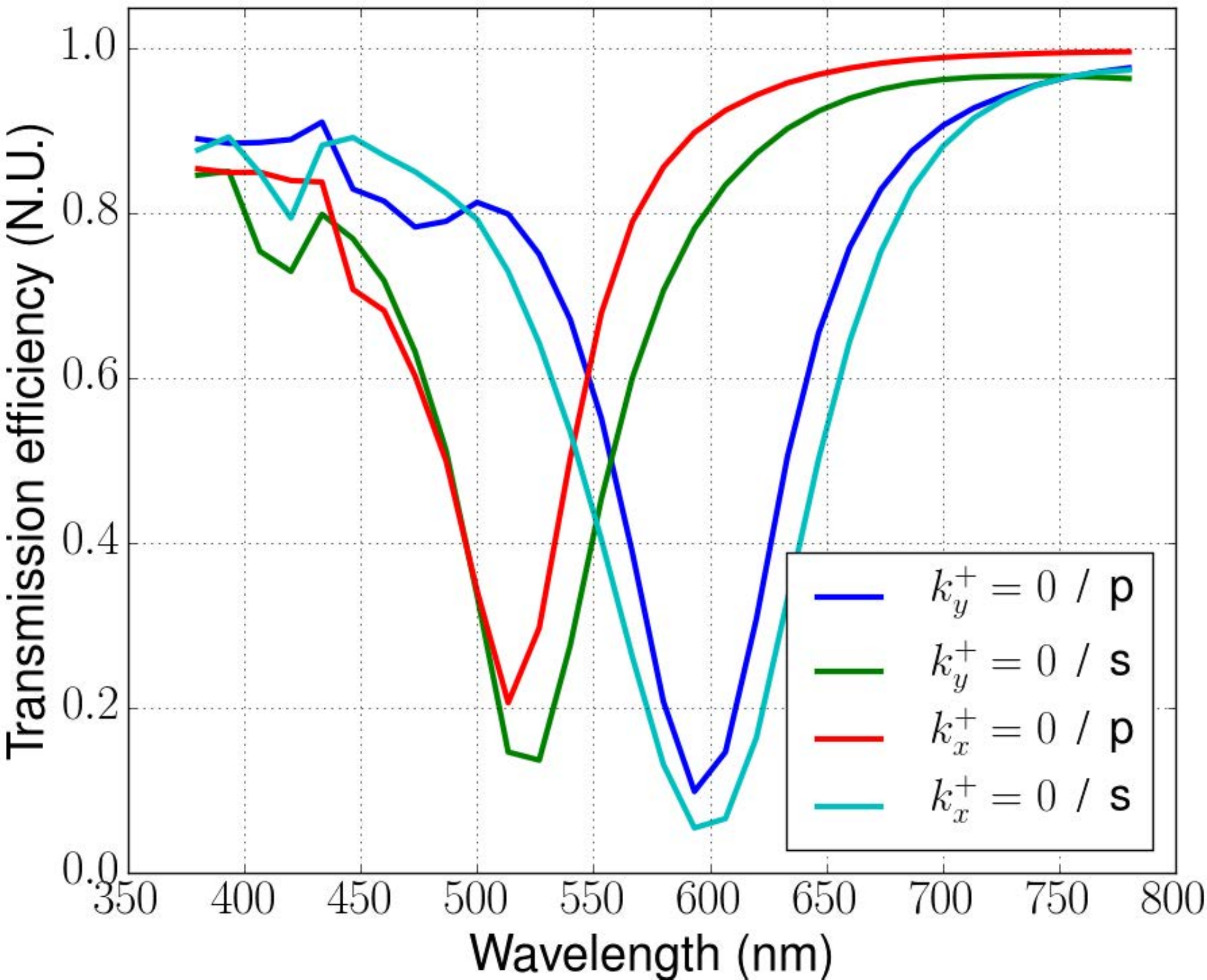}}
\caption{Reflexion (left panel) and transmission (right panel) 
efficiencies in the specular order for the structure F1.}
\label{fig:ROpti590}
\end{figure}
This geometry leads to reflexion peaks in the specular order centered
at two different wavelengths (525~nm and 590~nm) depending on the incidence 
plane and polarization.

The second proposed filter F2 is made of nanocylinders with $40$~nm height (right panel of Fig. 
\ref{F1-F2}). The ellipse diameters 
are $d_x = 100$~nm and $d_y = 60$~nm. The grating 
is embedded into SiO$_2$ (cover layer of $200$~nm and substrate). 
The square periodicity $a$ of the grating remains $200$~nm, and the illumination conditions 
are the same than for F1.
\begin{figure}[h]
\centering
\fbox{\includegraphics[width=0.48\linewidth]{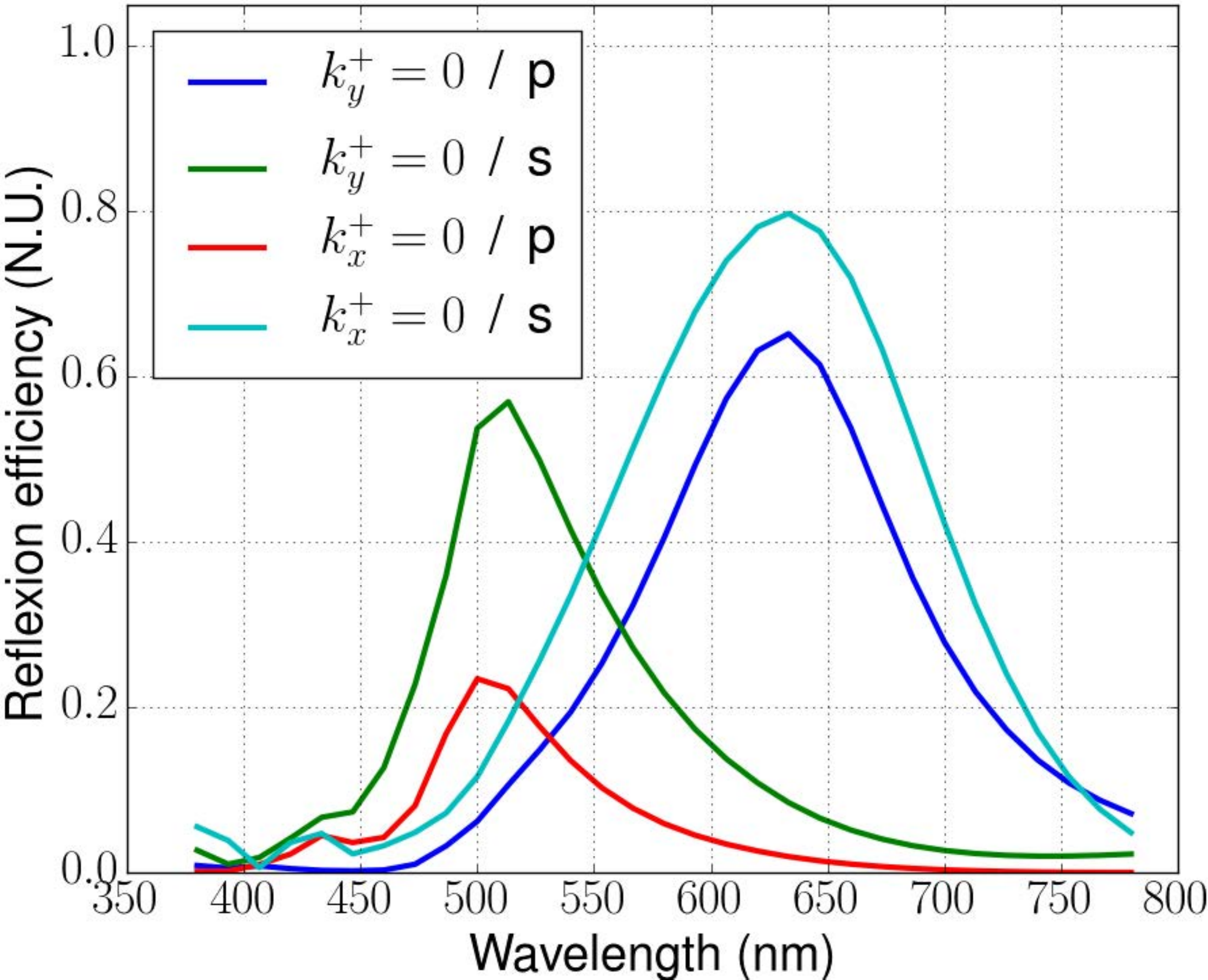}
\includegraphics[width=0.48\linewidth]{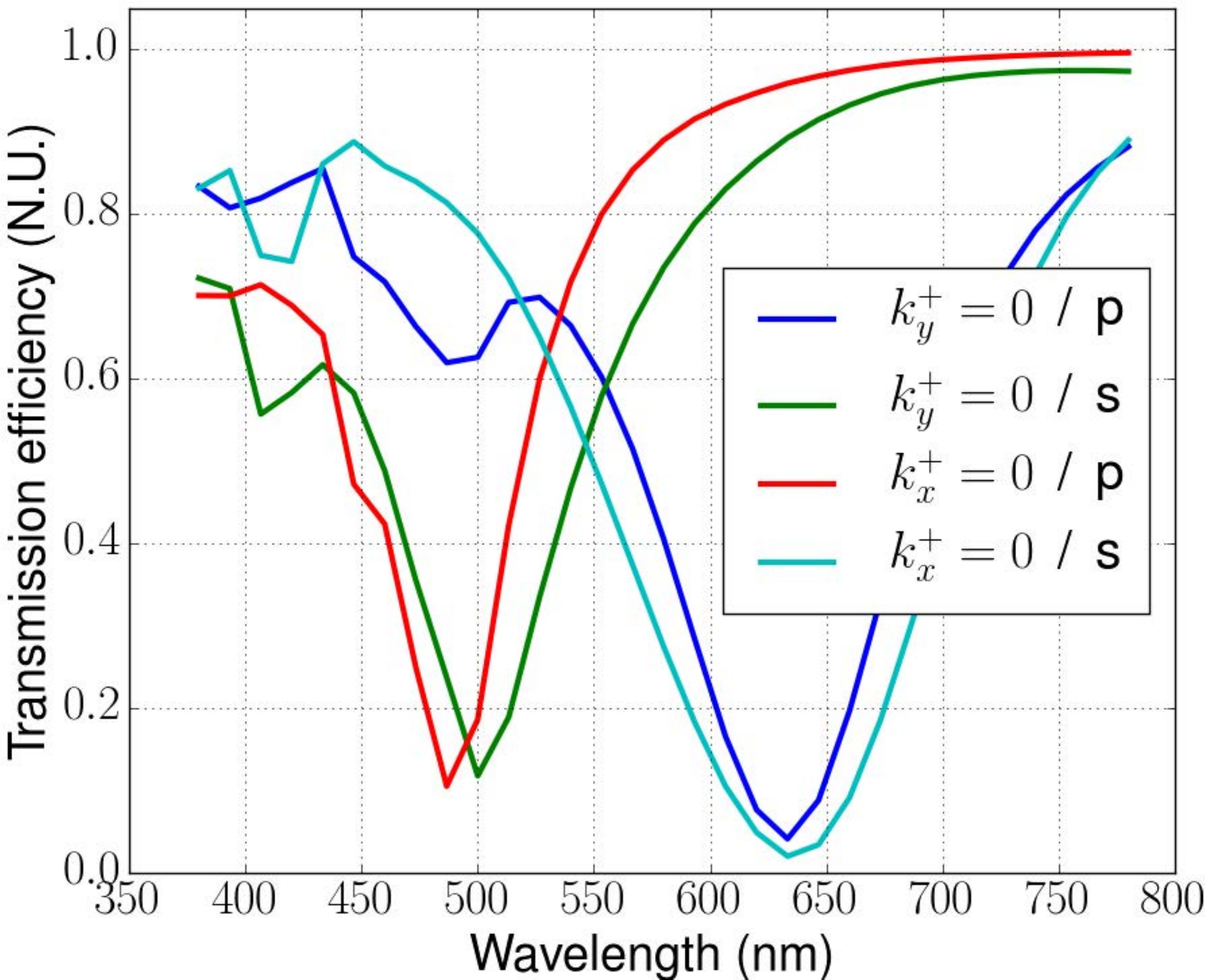}}
\caption{Reflexion (left panel) and transmission (right panel) efficiencies 
in the specular order for the structure F2.}
\label{fig:ROpti633}
\end{figure}
Figure \ref{fig:ROpti633} shows the different components of the energy balance for grating F2.
Here, reflexion peaks are obtained at the two different wavelengths 
$520$~nm and $633$~nm.
In both cases the averaged transmission over the whole visible spectrum is above 64\%, 
which ensures correct transparency. 
This is confirmed by the absorption spectra of the two structures reported on Fig. 
\ref{fig:AbsOpt}.

\begin{figure}[h]
\centering
\fbox{\includegraphics[width=0.48\linewidth]{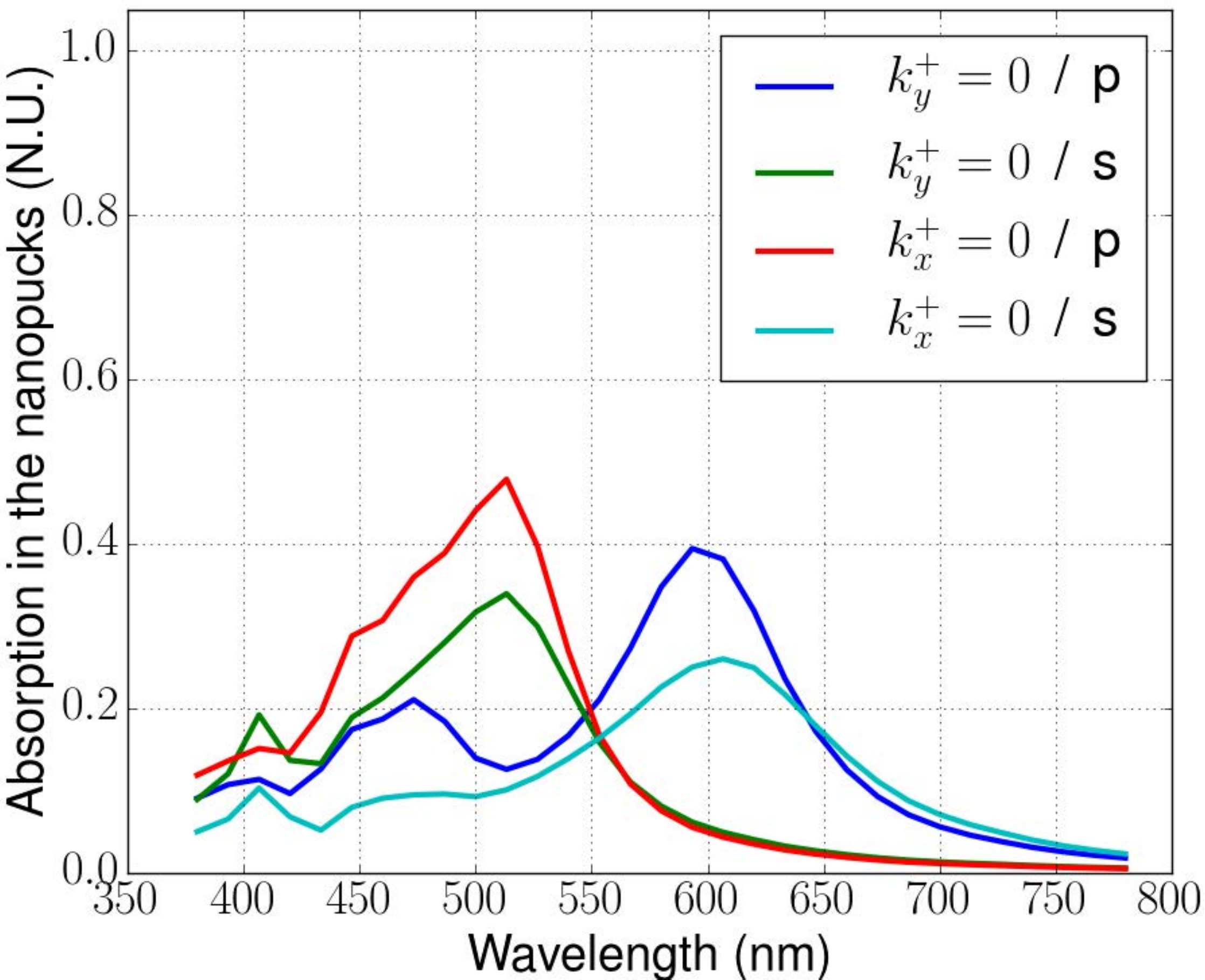}
\includegraphics[width=0.48\linewidth]{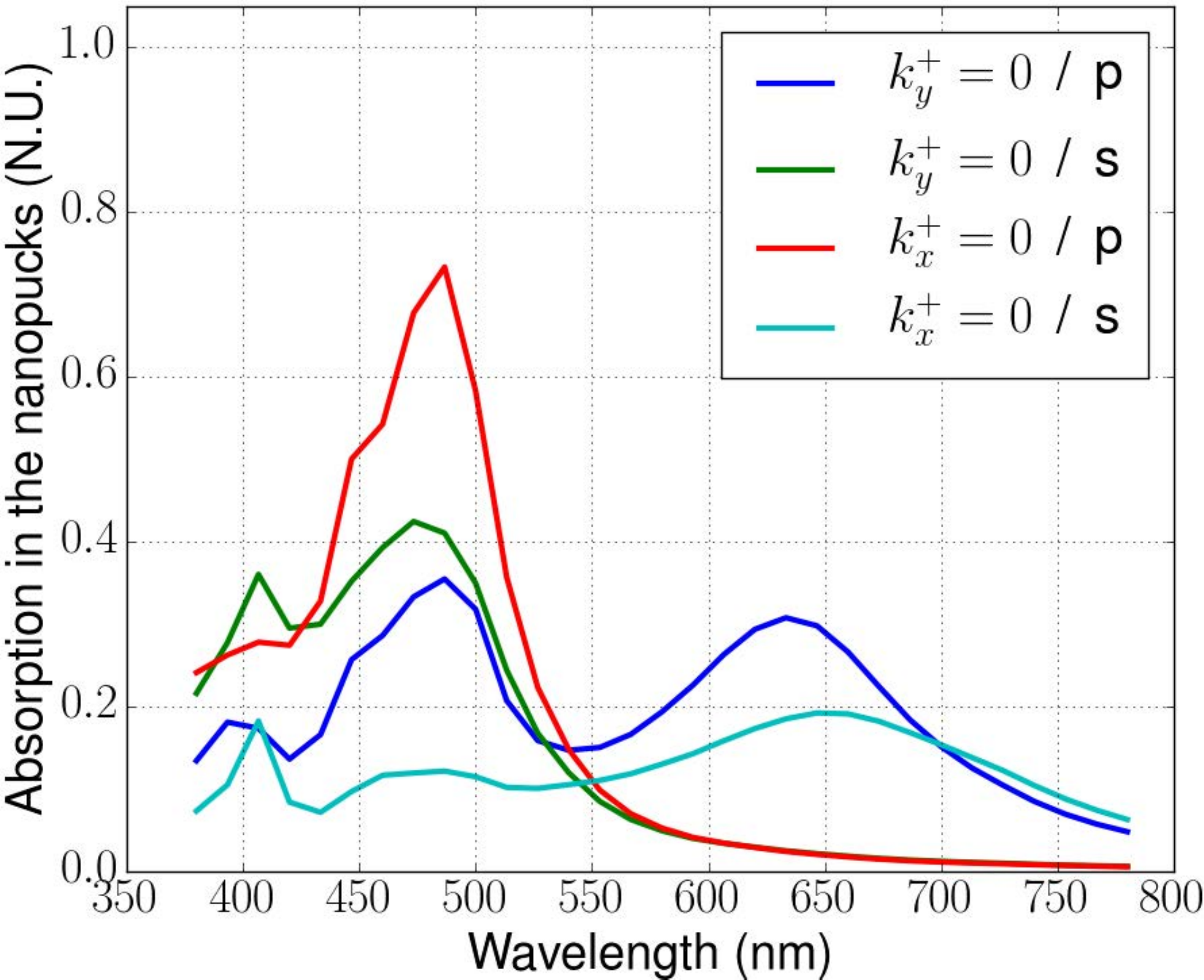}}
\caption{Absorption in the nanocylinders for structures F1 (left panel) and F2 
(right panel).}
\label{fig:AbsOpt}
\end{figure}

\section{Homogenization of nanocylinders structures}
\label{sec:Homog}
In this section, the properties of nanoparticles gratings are modeled using 
Maxwell-Garnett homogenization. 
{\color{black}{A detailed discussion on this homogenization of elliptic 
nanoparticles can be found in \cite{Hom96}, and more recently in 
\cite{Hom09}.
}}

In the present case, we consider anisotropic inclusions for which the 
effective permittivity components $\epsilon_j$ $(j= x,y,z)$ are given by:
\begin{equation}
\epsilon_j - \epsilon_m = \epsilon_m \, \dfrac{f(\epsilon_{p} - \epsilon_m)}{\epsilon_m + L_{j} (\epsilon_p - \epsilon_m)} 
\left[ 1 - L_{j}\, \dfrac{f (\epsilon_{p} - \epsilon_m)}{\epsilon_m +L_{j}(\epsilon_p -\epsilon_m)} \right]^{-1} \, .
\label{MG}
\end{equation}
Here, $\epsilon_p$ and $\epsilon_m$
are the permittivity of particles and surrounding medium respectively, $f$ is the filling fraction of metal, 
and $L_{j}$ are the components of the depolarization dyadic in the corresponding axis $(j= x,y,z)$.
We consider a cylindrical inclusion of elliptical cross section oriented in the z-direction embedded into a 
rectangular box. The height of the rectangular box is equal to the height $2h$ of the cylinder 
(see Fig. \ref{fig:UnitCell}). The basis of the rectangular box is the square unit cell of lattice.
Hence the filling fraction $f$ is just the ratio of the cylinder section 
(an ellipse of radius $R_x = d_x/2$ and $R_y = d_y/2$)
with respect to $a^2$, \textit{i.e.} $f = \pi R_x R_y/a^2$.
\begin{figure}[h]
\centering
\fbox{\includegraphics[width=0.5\linewidth]{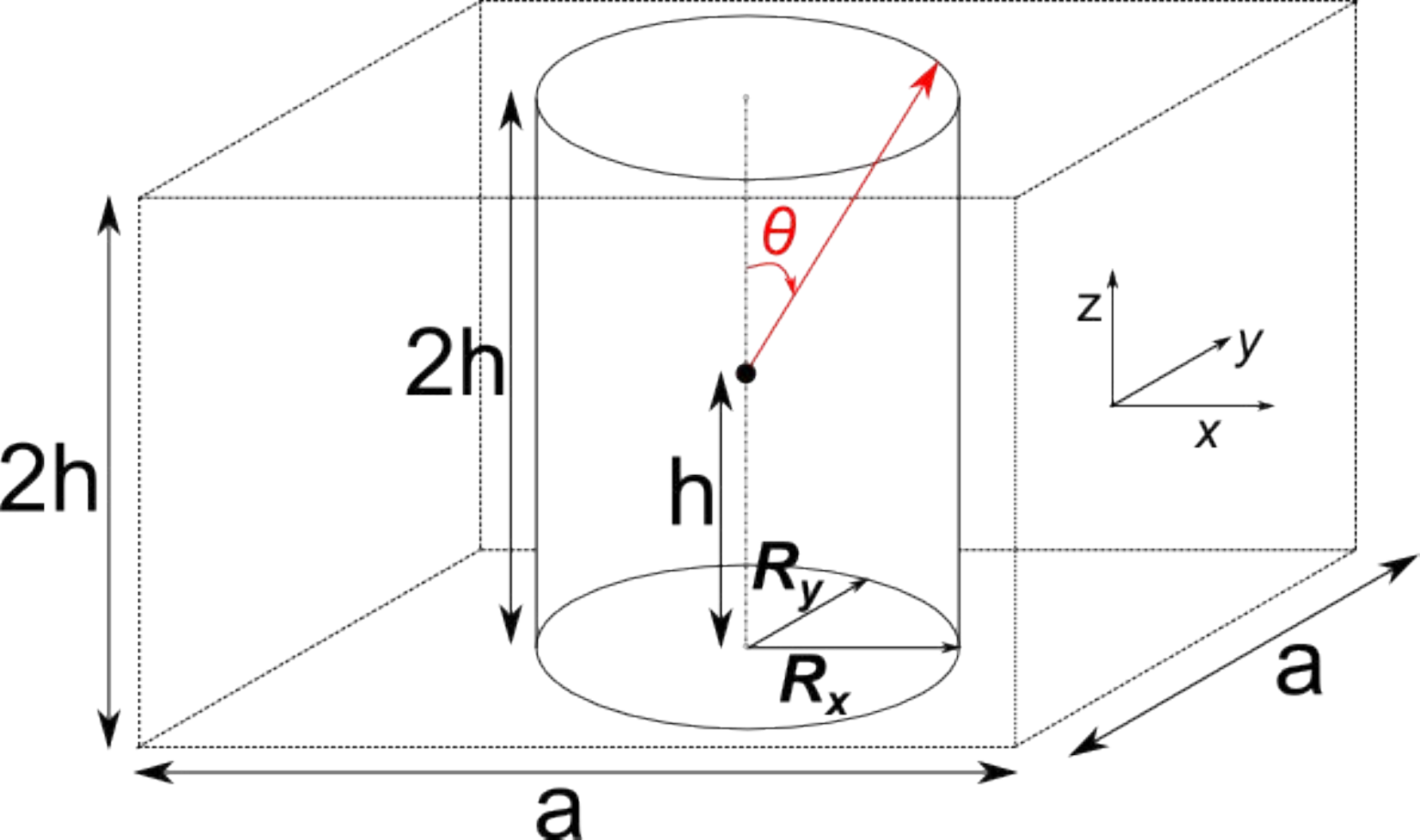}}
\caption{Scheme of the considered unit cell.}
\label{fig:UnitCell}
\end{figure}

The calculation of the depolarization components is reported in the appendix, in the 
both cases of cylinders with circular (appendix A) and elliptic (appendix B) cross section.
After computing the different depolarization factors, the effective permittivity 
components are calculated from Eq. (\ref{MG}).

This effective model has been applied to a square unit cell ($a=200$~nm) composed 
of a silver cylinder of elliptical cross section with radii $R_x = 40$~nm 
and $R_y = 30$~nm and height $2h$=30~nm.
The considered background medium is SiO$_2$. Hence, this unit cell represents the one of the 
grating corresponding to the structure F1.
The following depolarization components are obtained using Eqs. (\ref{Lz},\ref{LxLy}): 
$L_x=0.16$, $L_y=0.24$ and $L_z=0.60$. The corresponding effective anisotropic relative 
permittivity components are reported in left panel Fig. \ref{fig:EpsF1}. One can check that the 
effective permittivity components show resonances around 480~nm and 575~nm. 
\begin{figure}[h]
\centering
\fbox{\includegraphics[width=0.48\linewidth]{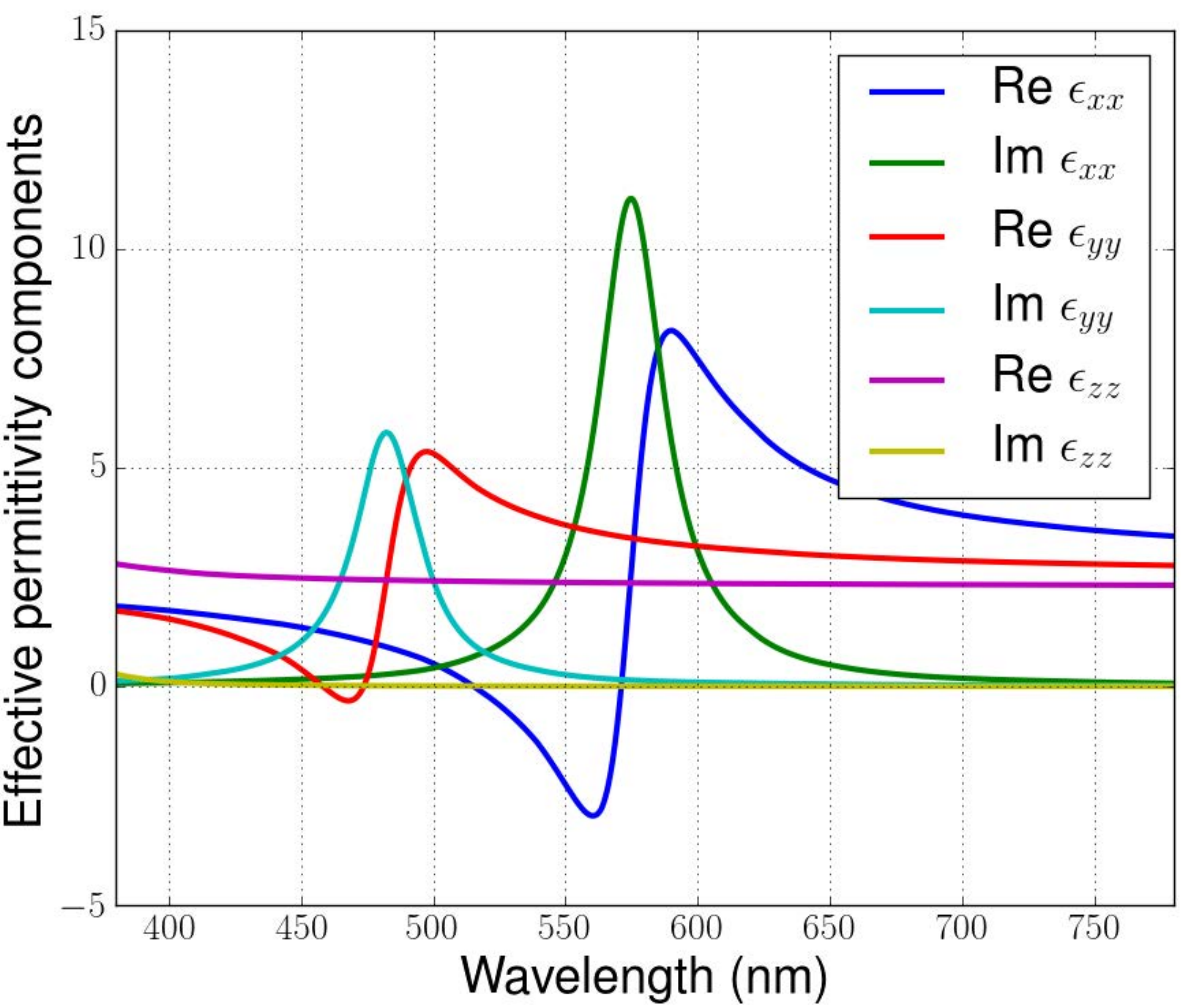}
\includegraphics[width=0.48\linewidth]{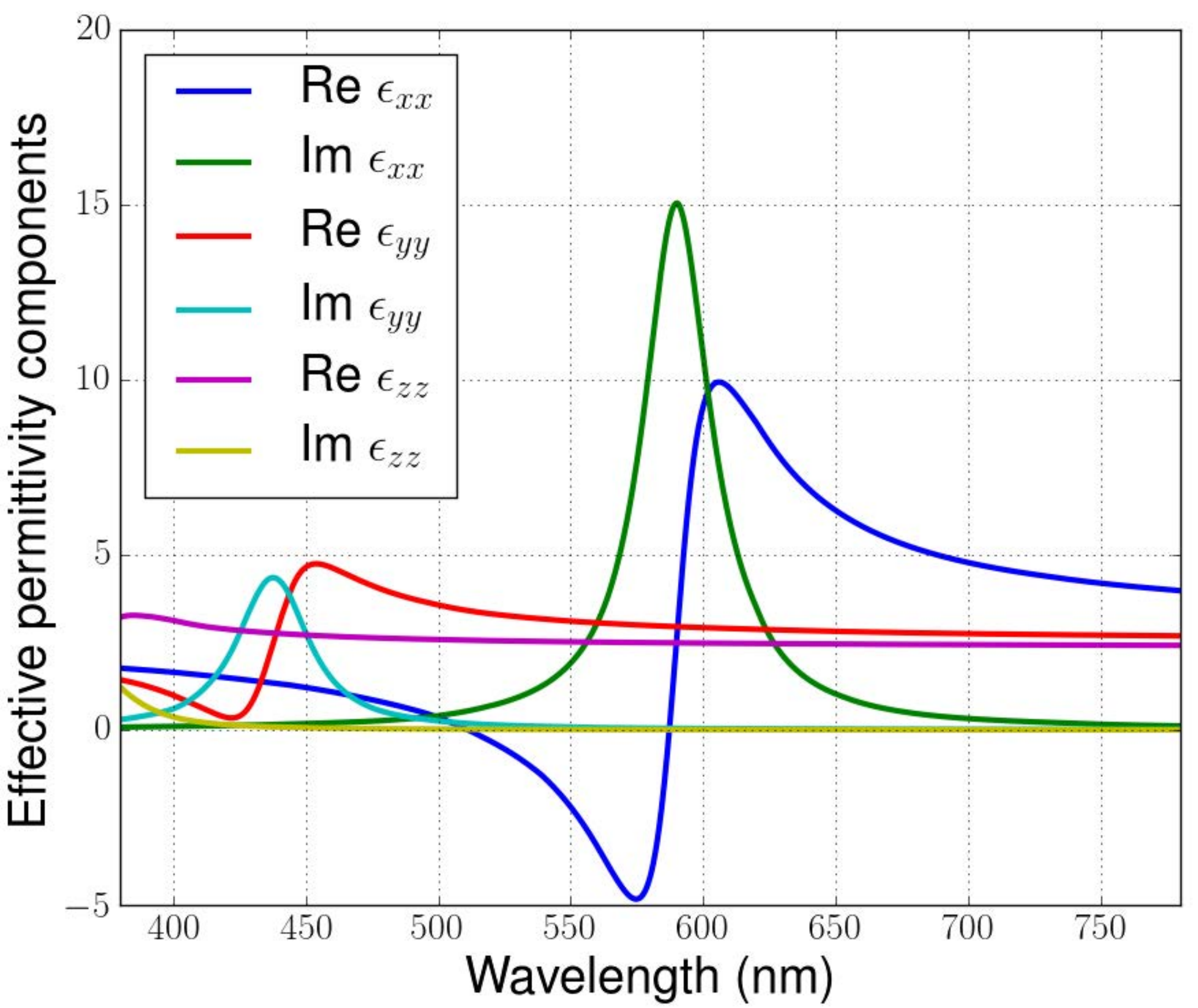}}
\caption{Effective anisotropic relative permittivity components for the unit cell 
of structures F1 (left panel) and F2 (right panel).}
\label{fig:EpsF1}
\end{figure}
Next the multilayer corresponding to the structure F1 
is considered. The multilayer is composed of (from top to bottom): 1) an incident medium (air); 2) a 
200~nm SiO$_2$ layer corresponding to the cover layer; 3) a 30~nm homogenized layer of effective
 permittivity given in left panel of Fig. \ref{fig:EpsF1} corresponding to the grating; 4)
a SiO$_2$ substrate. The reflexion and transmission coefficients 
have been calculated for an incident plane wave with incident wavevector that
lies either in the $xz$-plane ($k_y = 0$) either in the $yz$-plane ($k_x = 0$), 
with an angle of incidence fixed to $\theta_0 = 45^\circ$, 
and for wavelengths spanning the visible range.
Both p and s polarizations have been considered and compared to the rigorous 
FEM calculation of subsection 3.\ref{sec:Opti}. This comparison is shown in 
Fig.\ref{fig:CompF1}.

\begin{figure}[h]
\centering
\fbox{\includegraphics[width=0.48\linewidth]{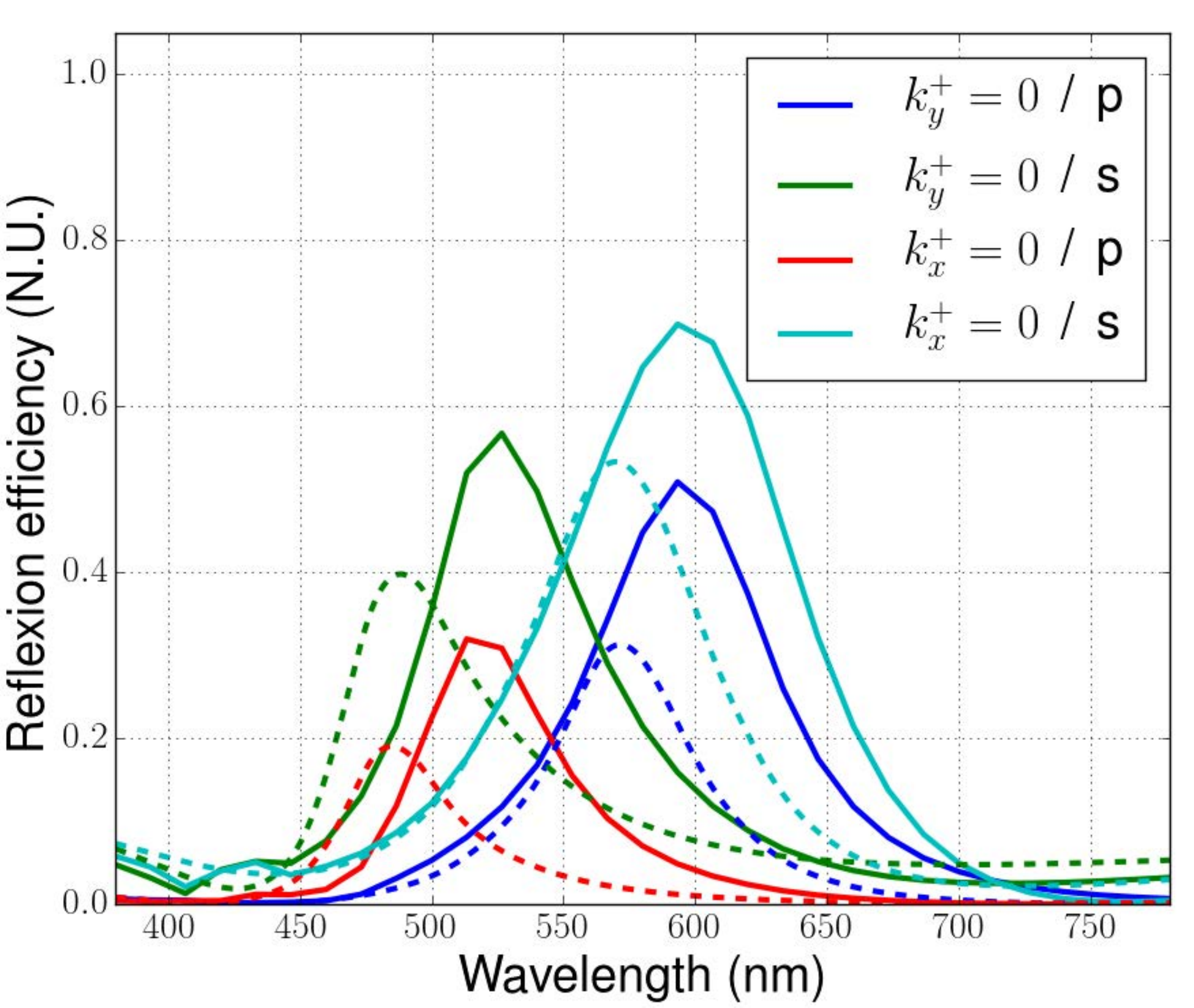}
\includegraphics[width=0.48\linewidth]{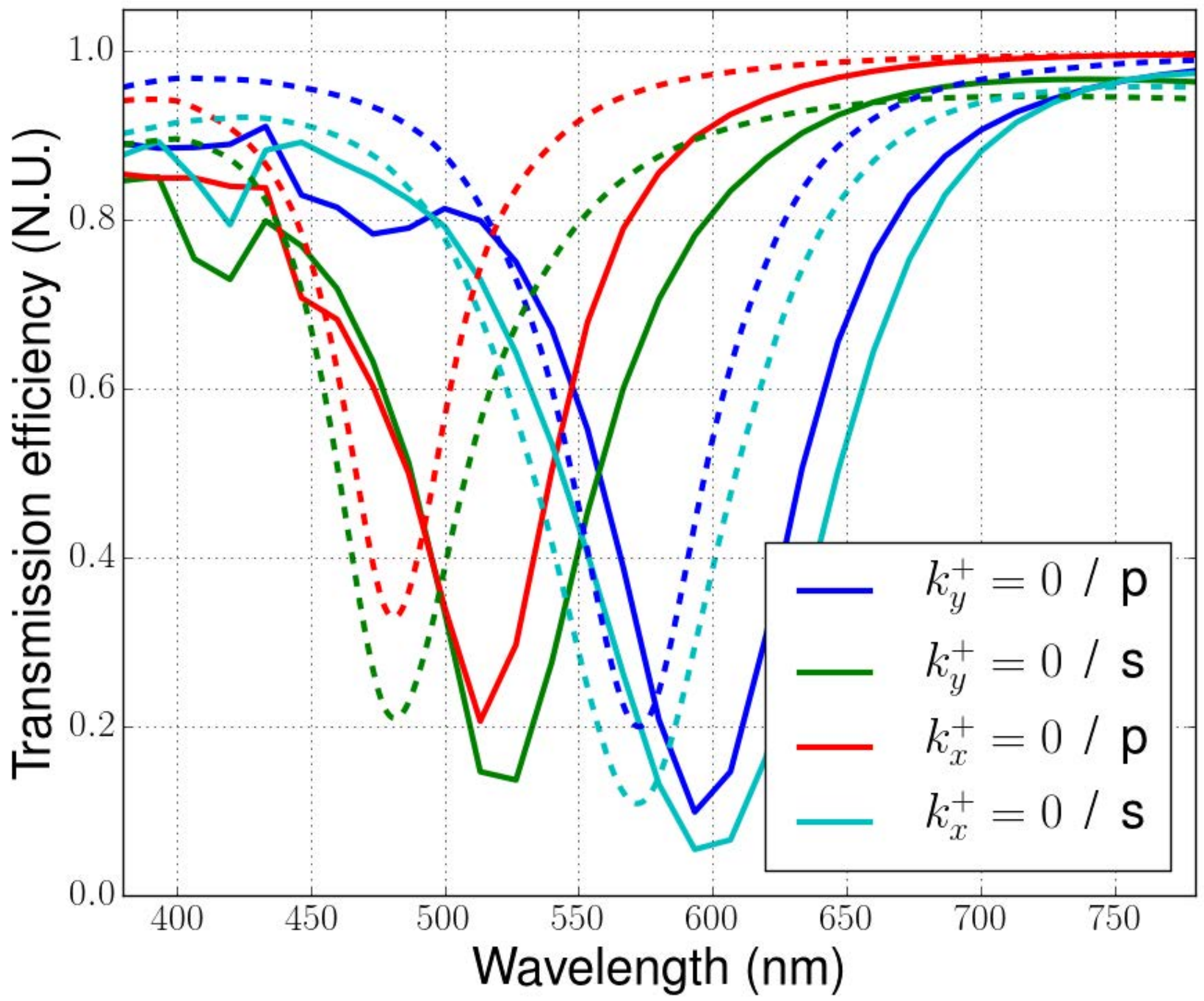}}
\caption{Reflexion (left panel) and transmission (right panel) efficiencies of structure F1 
obtained with the rigorous FEM calculation (line)
and with the homogenization model (dashed).}
\label{fig:CompF1}
\end{figure}

The same reasoning has been applied to the structure F2 and leads to the following depolarization 
components: $L_x=0.16$, $L_y=0.32$ and $L_z=0.52$. The corresponding effective relative 
permittivity is plotted in right panel of Fig.\ref{fig:EpsF1}, and presents resonances 
around 440~nm and 575~nm. The reflexion and transmission coefficients for the effective 
multilayer are compared to FEM results in Fig.\ref{fig:CompF2}.

One can observe that the homogenization leads to resonances which are in relative agreement 
with the rigorous FEM results, except a shift about 50~nm for the resonance wavelength. 
This shift can be attributed to the effect of the finite size of particles, since it can be checked 
numerically that this shift decreases together with the particle size. 
These resonances are both present in the effective permittivity 
components and the spectra of transmission and reflexion. 
Hence these homogenization results provide useful tendencies with much faster computation 
time, and can be used to design filtering properties. 

\begin{figure}[h]
\centering
\fbox{\includegraphics[width=0.48\linewidth]{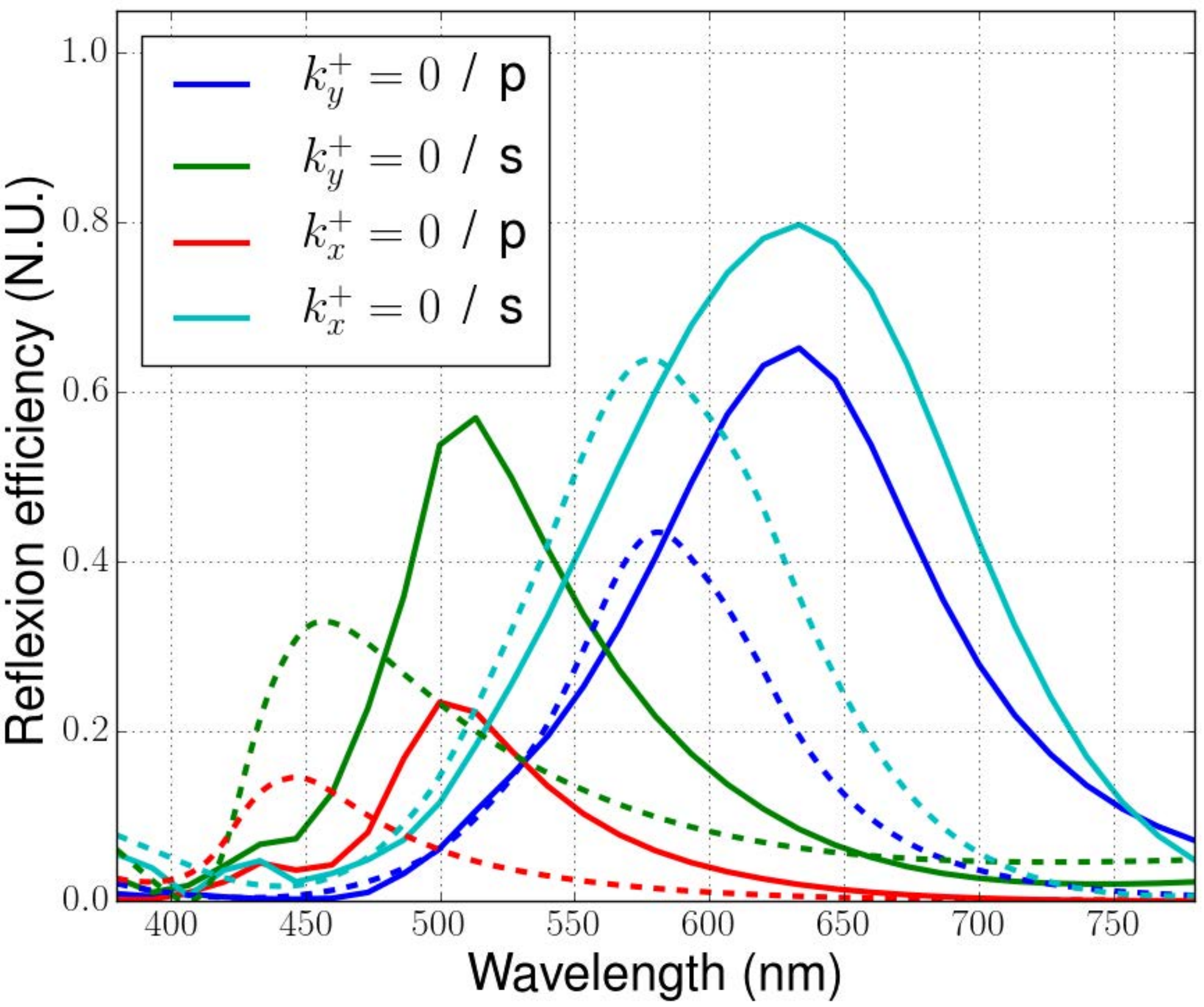}
\includegraphics[width=0.48\linewidth]{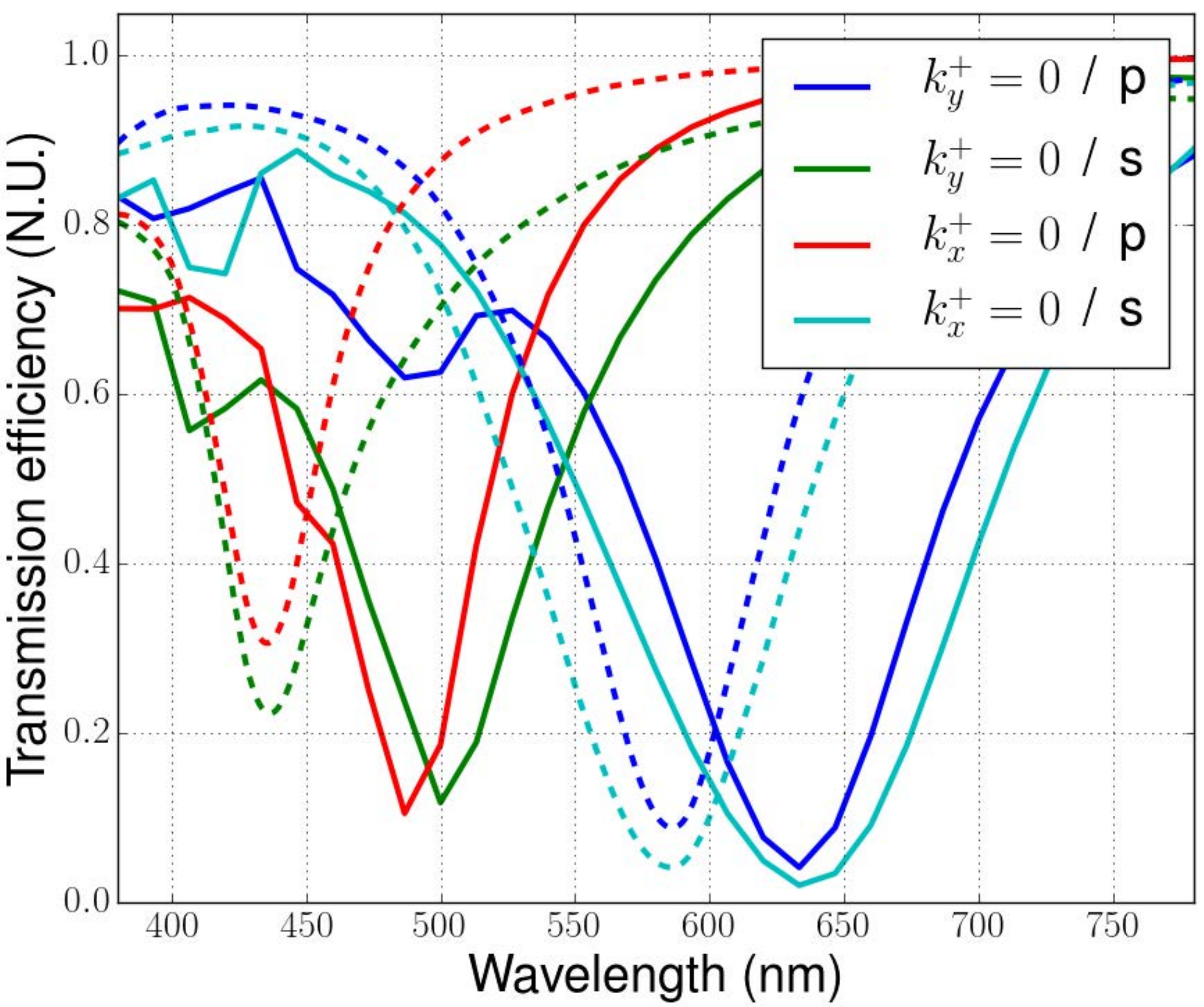}}
\caption{Reflexion (left panel) and transmission (right panel) efficiencies of structure F2 obtained 
with the rigorous FEM calculation (line) and with the homogenization model (dashed).}
\label{fig:CompF2}
\end{figure}

Notice that a similar homogenization model can be derived in the more complex situation were 
the particles stand directly on a substrate, without cover layer 
(for instance structures S5 or S6). 
For such case, the calculation of the depolarization factors is given in the 
appendix C.

\section{Diffraction by a single nanocyclinder}
\label{sec:Guillaume}
  \guill  
\def\bEinc{\boldsymbol{E}_0}
\def\bEd{\boldsymbol{E}_d}
\def\bEtot{\boldsymbol{E}}
  In this last section, the contribution of the single nanoparticle to the electromagnetic behavior 
  of the grating is analyzed. To that end, the problem of plane wave scattering by a single nanoparticle 
  is solved numerically using the FEM via an approach similar to Sec.~\ref{sec:Numer}.A
  changing Bloch-Floquet boundary conditions 
  to perfectly matched layers. Finally, a multipole expansion of the diffracted field is performed and discussed.
  
  Consider a non magnetic scatterer of arbitrary shape, relative permittivity $\varepsilon_s^b$
  embedded in a transparent non magnetic homogeneous background of relative permittivity 
  $\varepsilon_r^b$. We denote by $\varepsilon_r\ofx$ the piecewise 
  constant function equal to 
  $\varepsilon_s^b$ within the scatterer and $\varepsilon_r^b$ elsewhere. The scatter is 
  enlighten by a plane wave of arbitrary incidence and polarization $\bEinc$. 
  This incident field, i.e. the field in absence of the scatterer is the solution of
  the vector Helmholtz propagation equation in the resulting homogeneous medium:
  \begin{equation}\label{eq:helmholtz_pw_hom}
    \rot \rot \bEinc\ofx + k_0^2\,\varepsilon_r^b\,\bEinc\ofx = \mathbf{0}\,.
  \end{equation}
  We are looking for the total field $\bEtot$, resulting from the electromagnetic interaction of 
  the scatterer and the incident field, which is the solution of the Helmholtz equation:
  \begin{equation}\label{eq:helmholtz_tot_hom}
    \rot \rot \bEtot\ofx + k_0^2\,\varepsilon_r\ofx\,\bEtot\ofx = \mathbf{0}\,.
  \end{equation}
  The field diffracted or scattered by the object is defined as $\bEd:=\bEtot-\bEinc$.
  Combining Eqs.~(\ref{eq:helmholtz_pw_hom}, \ref{eq:helmholtz_tot_hom}) 
  allows to reformulate scattering problem to a 
  radiation problem. The scattered field now $\bEd$ satisfies the following propagation equation :
  \begin{equation}\label{eq:helmholtz_hom}
    \rot \rot \bEd\ofx + k_0^2\,\epsr\ofx\,\bEd\ofx = k_0^2\,(\epsr\ofx-\varepsilon_r^b)\,\bEinc\ofx\,
  \end{equation}
  such that $\bEd$ satisfies a radiation condition. The right hand side can be viewed as 
  a volume (current) source term with support the scatterer itself, $\epsr\ofx-\varepsilon_r^b$ being equal to 
  $\varepsilon_s^b-\varepsilon_r^b$ inside the scatterer and 0 elsewhere.
  
  The scattered field $\bEd$ can be expanded on the basis of outgoing vector partial
  waves $(\Mnm,\Nnm)$ built upon the vector spherical harmonics $(\Xnm,\Ynm,\Znm)$ following the framework and 
  conventions described in detail in \cite{Stout06I,Stout06II}:
      \begin{eqnarray}
       \Mnm\ofkr &=& h_n^{(+)}(kr)\,\Xnm\oftp \\
       \Nnm\ofkr &=&  \frac{1}{kr} \displaystyle \left[ \sqrt{n(n+1)}\,h_n^{(+)}(kr)\,\Ynm\oftp \right. \\
                 &+&   \xi'(kr)\,\Znm\oftp \displaystyle \bigg] \nonumber
      \end{eqnarray}
  The so-called multipole expansion is finally given by :
  \begin{equation}\label{eq:expansion}
    \bEd\ofr = \displaystyle\sum_{n=1}^{n_{\mathsf{max}}}\sum_{m=-n}^{m} \fhnm\,\Mnm\ofkr+\fenm\,\Nnm\ofkr\,\,
  \end{equation}
  where $\fhnm$ and $\fenm$ can be numerically computed on any sphere of radius $R$ englobing the scatterer :
  \begin{eqnarray}\label{eq:coeff_expansion}
    \fenm&\hspace*{-3mm}=&\hspace*{-3mm}\frac{kR}{\xi'(kR)} \displaystyle\int_0^{2\pi}\int_0^{\pi} \bEd(R,\theta,\varphi)\cdot\Znm^\star(\theta,\varphi)\,\mathsf{d}\theta\,\mathsf{d}\varphi \\
    \label{eq:coeff_expansion2}
    \fhnm&\hspace*{-3mm}=&\hspace*{-3mm}\frac{kR}{h_n^{(+)}(kR)} \displaystyle\int_0^{2\pi}\int_0^{\pi} \bEd(R,\theta,\varphi)\cdot\Xnm^\star(\theta,\varphi)\,\mathsf{d}\theta\,\mathsf{d}\varphi
  \end{eqnarray}
  Note that $\bEd$ in Eqs.~(\ref{eq:coeff_expansion}) and (\ref{eq:coeff_expansion2}) is expressed from 
  the FEM inherited cartesian coordinates to spherical coordinates.
  This expansion is the classical way to obtain the far field radiation pattern of antennas.
  It would require major adjustments to take into account the substrate and superstrate. 
  First of all, let us state that the resonant phenomena described in the previous section do not depend on
  the close vicinity of the interface. Figure~\ref{fig:conf_opti_in_hom} shows the reflectivity of 
  silver nanocylinders grating of the optimized structure F1, 
  when embedded in an homogeneous infinite $\mathrm{SiO}_2$ background. In this simplified model, 
  the $\theta_0$ incidence is adjusted to compensate for the missing refraction at $\mathrm{SiO}_2$ interface 
  $\theta_0=\mathrm{arcsin}(\mathrm{sin}(45^\circ)/n_{\mathrm{SiO}_2}) \approx29^\circ$. The same four resonances as in 
  Fig.~\ref{fig:AbsOpt} are obviously excited with $\theta=29^\circ$ 
  depending on the plane of incidence (compare dashed and plain lines in Fig.~\ref{fig:conf_opti_in_hom}). The only noticeable
  difference in s-polarization case (see green and cyan lines in Fig.~\ref{fig:conf_opti_in_hom}) can be 
  partly explained by the fact that at $45^\circ$ incidence, reflexion on a bare silica diopter is $\sim 8.3\%$ in s-polarization 
  (while only $\sim 0.5\%$ in p-polarization, due to the Brewster effect).
  
  \begin{figure}[h]
  \centering
    \includegraphics[width=0.98\linewidth]{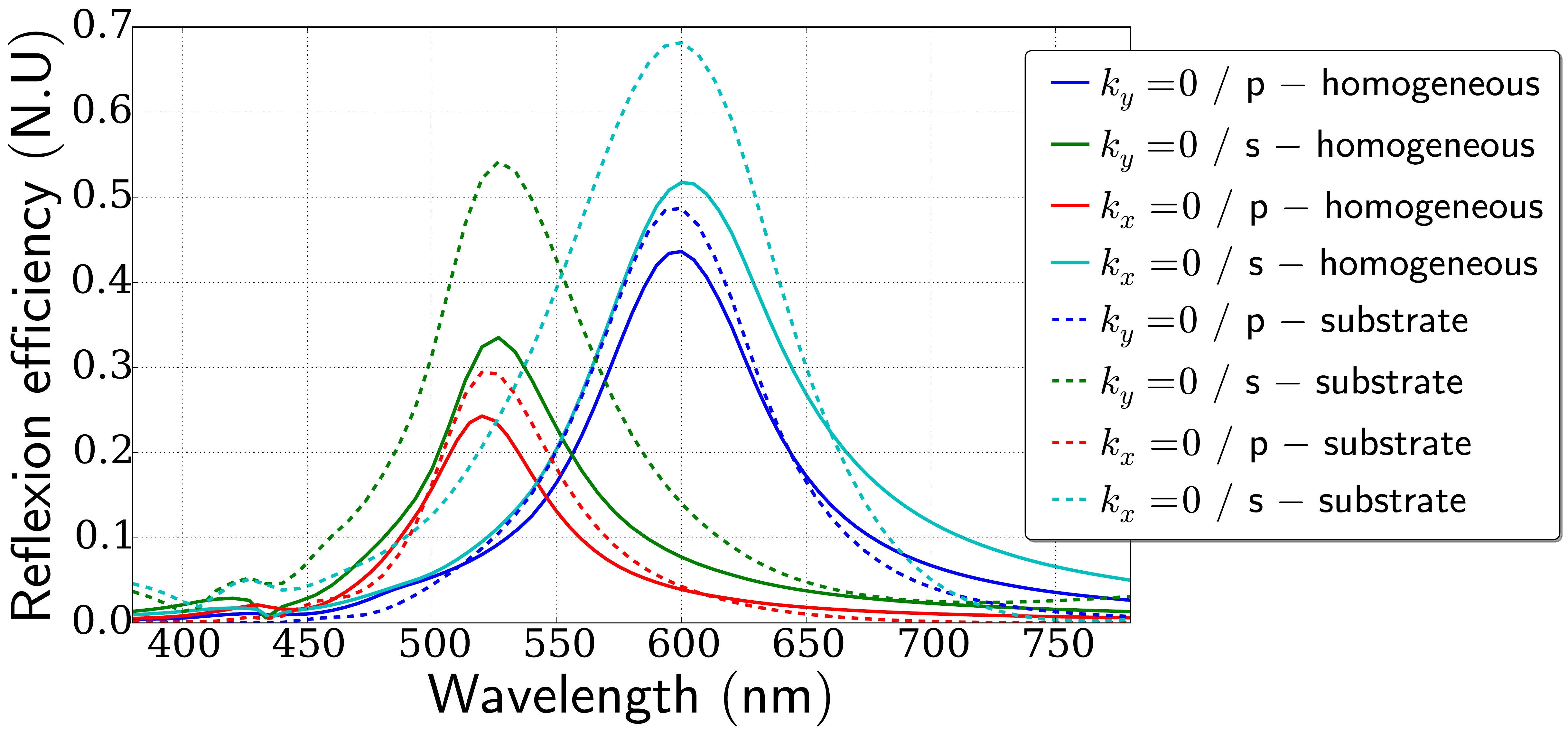}
    \caption{(Dashed lines) Reflexion efficiency in the specular order for the structure F1 (same as left panel of Fig. \ref{fig:ROpti590}). (Plain lines) Grating of silver nanocylinders of same dimensions as in optimized structure F1, 
    but embedded in homogeneous SiO$_2$ background. 
    Reflexion peaks in both panels are very similar.
    }
    \label{fig:conf_opti_in_hom}
  \end{figure}
  
  Consider now the single scatterer case. 
  First of all, the only non negligible terms in expansion Eq.~(\ref{eq:expansion})
  correspond to electric dipoles whatever the considered incidence.
  All higher electric orders and all magnetic orders are at least 50 fold lower in magnitude than the dipolar electric ones
  over the whole visible spectrum.
  As shown in Fig.~\ref{fig:multipoles}(a)~and~\ref{fig:multipoles}(b) for $k_y=0$, i.e. 
  when the plane of incidence contains the $r_x$-axis of the ellipse, two resonant scenarii occur  
  depending on the incident polarization. With p-polarization
  [Fig.~\ref{fig:multipoles}(a)], the incident electric field lies
  within the plane of incidence, and two electric multipoles dominate the far field scattered power, corresponding to 
  induced electric dipoles of moments along $Ox$ (see $f^{(e)}_{1,\pm1}$) and $Oz$ (see $f^{(e)}_{1,0}$).
  Heuristically, the incident electric field indeed only \emph{sees} two characteristic dimensions 
  of the scatterer, {i.e.} the height $h$ of the elliptic 
  nanocylinder and its greater diameter $2r_x$. 
  With $s$-polarization, the only induced dipole present is along $Oy$, corresponding to the fact that the electric 
  field now only \emph{sees} the smaller diameter $2r_y$. 
  The larger radius [Fig.~\ref{fig:multipoles}(a)] leads to a redshifted resonant response of the particle, 
  as already observed for gold circular cylinders (see Fig.~\ref{fig:Reflexion}).
  The same considerations hold for $k_x=0$  as depicted in Fig.~\ref{fig:multipoles}(c)~and~\ref{fig:multipoles}(d).
  Electric field maps of the periodic and isolated cases are very similar in spite of the rather small 200~nm bi-period,
  which further confirms the weak coupling between the elliptic nanocylinders. Above 450~nm, the subwavelength 
  bi-periodicity selects only the specular propagation direction among all by constructive interferences.
  \begin{figure}[h]
  \centering
    \includegraphics[width=0.98\linewidth]{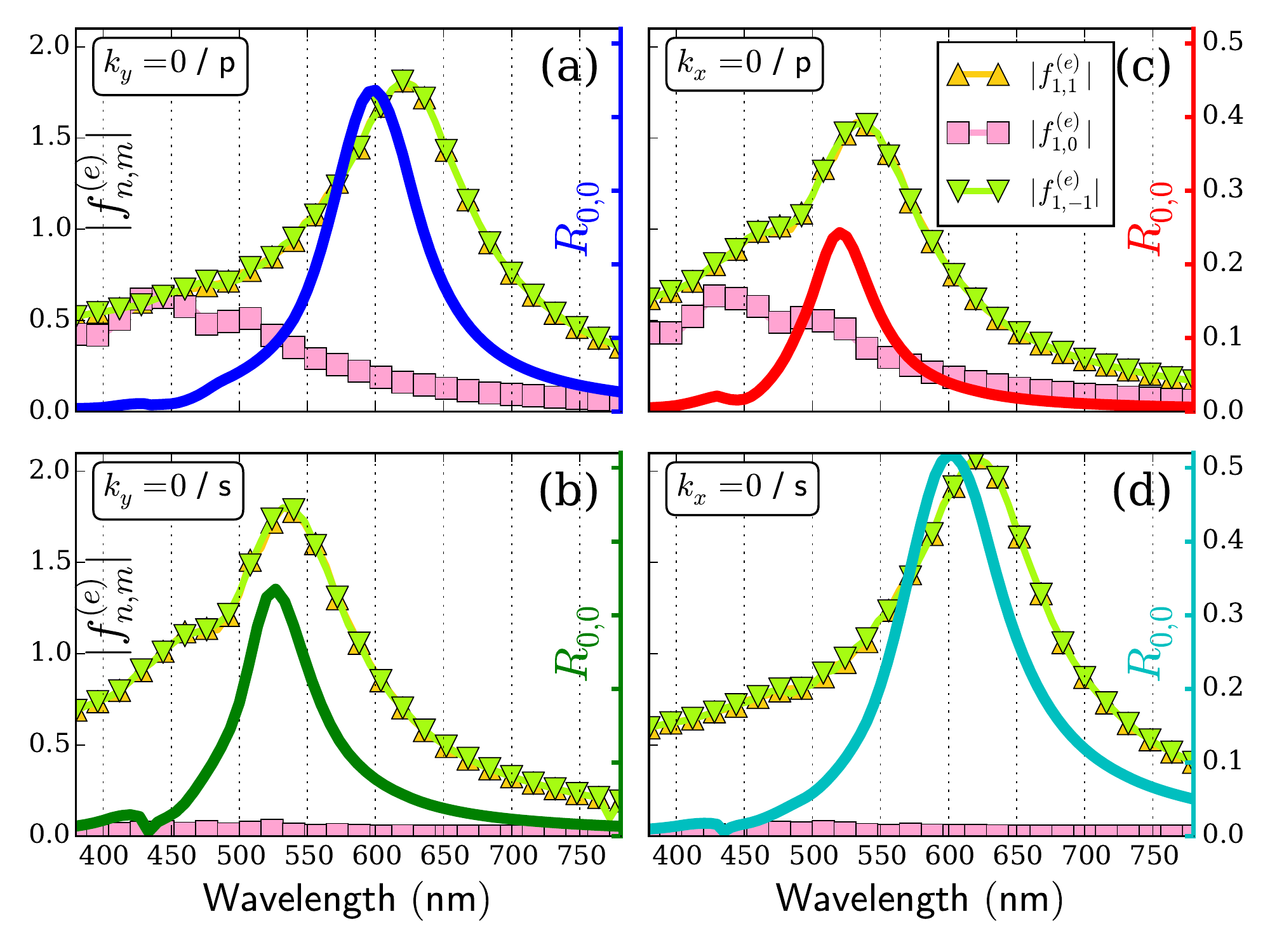}
    \caption{reflexion of the periodic structure compared to
     multipolar coefficients of the isolated scatterer for incidence 
     $\theta_0=29^\circ$, with (a) $k_y=0$  
     and p-polarization, (b) $k_x=0$  and 
     p-polarization , (c) 
     $k_y=0$ and s-polarization , (d) $k_x=0$  and s-polarization. }
    \label{fig:multipoles}
  \end{figure}
  \black

\section{Conclusions}
\label{sec:CCL}
We have shown the possibility to design efficient reflective filters with selective
wavelength mirror properties together with global transparency over the visible range. 
A numerical tool based on the FEM has been used to analyze the behavior of 
nanoparticles gratings, and comparison between numerical and experimental results has 
been performed. 
Two optimized filters with two operating wavelengths depending on the polarization 
have been proposed. It has been shown that Maxwell-Garnett homogenization 
can be used to model the considered structures. Also, the calculation of the radiation 
by a single nanoparticle has shown that the diffractive properties of the considered systems 
are governed by a single isolated particle. 

Whereas the main principle has been demonstrated, the full structure can be improved 
by several manners in order to increase the reflectivity and transmission. 
Especially, fabrication can be done now without ITO conductive layer and 
without Ti attaching layer. This should significantly reduce absorption. 
Besides, since the diffractive properties are governed by a single isolated particle, 
the same properties can be obtained with periodic or aperiodic structures as along as the 
size and the density of the nanoparticles are the same. This may also considerably ease 
the fabrication on large surfaces. 

These promising results pave the way of smart windows and could find some innovative 
applications in transport industry such as automotive or aeronautic sectors for 
displays, human-machine interfaces and sensors.

\section*{Funding Information}
This work has been supported by the ANR funded project PLANISSIMO (ANR-12-NANO-0003) 
and by the french RENATECH network.

%




\section{appendix}
\label{sec:appendix}
\subsection{Depolarization dyadic for circular cylinders}
The calculation of depolarization dyadic components associated with a finite circular 
cylinder is recalled. Let $V$ be the volume of a cylinder of height $2h$ 
and radius $R$, and let $\r$ be the vector of coordinates $(x,y,z)$ with the 
origin $\r = 0$ at the center of the cylinder. Then, the dimensionless 
depolarization dyadic at the center of the cylinder is formally given by
\begin{equation}
L = - \dint_V d \r \, \grad \grad \dfrac{1}{4 \pi |\r|} \, .
\end{equation}
This expression has to be considered carefully since a singularity is present 
at the origin. This difficulty is fixed by splitting the integral above into 
a first integral over a small sphere $B_0$ and a second integral over the 
remaining volume $V \setminus B_0$. Next, using that the integral over $B_0$
is just $1/3$ of the unit dyadic $U$, we have 
\begin{equation}
L = \dfrac{1}{3} U - \dint_{V \setminus B_0} d \r \, \grad 
\grad \dfrac{1}{4 \pi |\r|} \, .
\end{equation}
Now, the Green-Ostrogradski theorem can be applied. Let $S$ and $S_0$ be the 
surfaces of the cylinder $V$ and the sphere $B_0$ respectively. 
The expression above becomes
\begin{equation}
L = \dfrac{1}{3} U - \dint_{S} d \s \, \n 
\grad \dfrac{1}{4 \pi |\r|} \, + \dint_{S_0} d \s \, \n 
\grad \dfrac{1}{4 \pi |\r|} \, .
\end{equation}
where $ds$ is the infinitesimal surface element and $\n$ is the outgoing normal 
of the surfaces. Since the last term is just $U/3$, the expression reduces to 
\begin{equation}
L = - \dint_{S} d \s \, \n 
\grad \dfrac{1}{4 \pi |\r|} = \dint_{S} d \s \, \n 
\dfrac{\r}{4 \pi |\r|^3} \, .
\label{eq:start}
\end{equation}
Finally, using the symmetries of the cylinder, it is found that the component 
$L_z \equiv L_{zz}$ of the dyadic is given by the integration over the two 
horizontal disks at the top and bottom of the cylinder:
\begin{equation}
L_{z} = 2 \dint_{0}^R dr \, \dint_{0}^{2 \pi} r d \phi \, 
\dfrac{h}{4 \pi (h^2 + r^2)^{3/2}} = 1 - \cos \theta \, .
\end{equation}
As to the components $L_x \equiv L_{xx}$ and $L_y \equiv L_{yy}$ 
($L_y = L_x$ by symmetry), they 
are provided by the integration over the vertical face of the cylinder:
\begin{equation}
L_{x} = \dint_{-h}^h dz \, \dint_{0}^{2 \pi} R d \phi \, 
\dfrac{R \cos^2 \phi}{4 \pi (z^2 + R^2)^{3/2}} = \dfrac{1}{2} \cos \theta \, .
\end{equation}
Hence the well known expressions \cite{yaghjian1980electric,van2007electromagnetic} 
have been retrieved. 

\subsection{Depolarization dyadic for elliptic cylinders}
\label{sec:EllipticalCylinder}
In this subsection, $V$ and $S$ are the volume and the surface respectively 
of a cylinder with elliptic cross section 
of height $2h$ and radii $R_x$ and $R_y$.
All the procedure performed in the circular case remains correct until 
equation (\ref{eq:start}). Similarly to the previous case, using 
the symmetries of the elliptic cylinder, it is found that the component $L_z$ 
is given by the integration over the two horizontal surfaces $S_e$ defined by 
\begin{equation}
S_e = \{ (x,y) \, | \, x^2/R_x^2 + y^2 /R_y^2 \le 1 \, \} \, .
\end{equation}
Then the depolarization component in the $z$ direction is 
\begin{equation}
L_{z} = 2 \dint_{S_e} dx \, dy \, 
\dfrac{h}{4 \pi (h^2 + x^2 + y^2)^{3/2}} \, ,
\label{Lz}
\end{equation}
which requires numerical integration to determine its value. 
Again using the symmetries of the cylinder, it is deduced that the 
depolarization factors $L_x$ and $L_y$ are provided by the integration 
over the vertical face of the cylinder. The vector $\r$ describing this 
surface is (the third component $z$ can be omitted)
\begin{equation}
\r = \left( \begin{array}{c} R_x \cos \phi \\ R_y \sin \phi \end{array} 
\right) \quad \Longrightarrow \quad \dfrac{d \r}{d \phi} = 
\left( \begin{array}{c} - R_x \sin \phi \\ R_y \cos \phi \end{array} 
\right) 
\end{equation}
and the unit normal $\n$, orthogonal to $d \r / d \phi$, is then 
\begin{equation}
\n = \dfrac{1}{\sqrt{R_x^2 \sin^2 \phi + R_y^2 \cos^2 \phi}} 
\left( \begin{array}{c} R_y \cos \phi \\ R_x \sin \phi 
\end{array} \right)  \, .
\end{equation}
Then the tensor product of the two vectors $\n$ and $\r$ is given by 
\begin{equation}
\n \r = \dfrac{1}{\sqrt{R_x^2 \sin^2 \phi + R_y^2 \cos^2 \phi}}
\left( \begin{array}{lr} \!\!\! R_x R_y \cos^2 \phi & R_x^2 \cos \phi \sin \phi \!\!\! \\ 
\!\!\! R_y^2 \cos \phi \sin \phi & R_x R_y \sin^2 \phi \!\!\!
\end{array} \right) \!.
\end{equation}
Finally, since $ds = dz d\phi \times | d\r/d\phi | $, the obtained expressions 
for the depolarization are: 
\begin{equation}
\begin{array}{l}
L_{x} = \dint_{-h}^h dz \, \dint_{0}^{2 \pi} d \phi \, 
\dfrac{R_x R_y \cos^2 \phi}{4 \pi (z^2 + R_x^2 \cos^2 \phi + R_y^2 \sin^2 \phi)^{3/2}} \, , \\[2mm]
L_{y} = \dint_{-h}^h dz \, \dint_{0}^{2 \pi} d \phi \, 
\dfrac{R_x R_y \sin^2 \phi}{4 \pi (z^2 + R_x^2 \cos^2 \phi + R_y^2 \sin^2 \phi)^{3/2}} \, .
\label{LxLy}
\end{array}
\end{equation}
Numerical integration is required to obtain the values of these components.
\subsection{Depolarization dyadic for circular cylinders on a substrate}
The case of a circular cylinder lying on a substrate is considered. The notations are the 
same as in the appendix A, with the origin $\r=0$ at the center of the cylinder of 
height $2h$. In addition, a substrate of relative permittivity 
$\epsilon^-$ is located in the half space $z < -h$ with its plane interface at 
the bottom circular face of the cylinder. Let $\epsilon(\r)$ be the relative 
permittivity defining the environment of the cylinder: $\epsilon(\r) = 1$ if 
$z > -h$ and $\epsilon(\r) = \epsilon^-$ if $z < -h$. 
The electrostatic potential $\Phi(\r)$ created by a point source is the 
solution of 
\begin{equation}
\grad \cdot \epsilon(\r) \grad \Phi(\r) = \delta(\r)
\label{eqPhi}
\end{equation}
Denoting $\r_0 = 2 h \boldsymbol{e}_z$, the solution of the equation 
above is 
\begin{equation}
\begin{array}{l}
z > -h \, : \quad \Phi(\r) = - \dfrac{1}{4 \pi |\r|} - 
\dfrac{1 - \epsilon^-}{1 + \epsilon^-} \dfrac{1}{4 \pi |\r +\r_0|} \, , \\[2mm]
z < -h \, : \quad \Phi(\r) = - 
\dfrac{2}{1 + \epsilon^-} \dfrac{1}{4 \pi |\r|} \, .
\end{array}
\end{equation}
The depolarization dyadic is then given by an expression equivalent 
to equation (\ref{eq:start}):
\begin{equation}
L = - \dint_{S} d \s \, \n 
\grad \Phi(\r) \, .
\label{eq:substrate}
\end{equation}
Next, the calculation is similar to the one in the appendix A, with an additional 
term corresponding to the reflexion part of $\Phi(\r)$. The resulting 
expression for the vertical component of the depolarization dyadic is
\begin{equation}
L_z = 1 - \cos \theta + \dfrac{1 - \epsilon^-}{1 + \epsilon^-}
\dfrac{\cos\theta - \cos\theta_1}{2} \, , 
\end{equation}
where
\begin{equation}
\cos \theta = \dfrac{h}{\sqrt{h^2 + R^2}} \, , \quad 
\cos \theta_1 = \dfrac{3 h}{\sqrt{(3h)^2 + R^2}} \, .
\end{equation}
The other components of the dyadic are
\begin{equation}
L_x = L_y = \dfrac{\cos \theta}{2} - \dfrac{1 - \epsilon^-}{1 + \epsilon^-}
\dfrac{\cos\theta - \cos\theta_1}{4} \, .
\end{equation}
This calculation can be extended to cylinders with elliptic cross section 
by following the procedure of appendix B.

\end{document}